\documentclass[letter,11pt]{article}

\usepackage{graphicx,amsmath,amsfonts,algorithm}
\floatname{algorithm}{Protocol}
\usepackage{graphics}
\usepackage{amssymb}
\usepackage{amsmath}
\usepackage{amsthm}
\usepackage{complexity}
\usepackage{color}
\usepackage{bbm}
\usepackage[affil-it]{authblk}
\usepackage[margin=1in]{geometry}
\usepackage[normalem]{ulem}
\usepackage{makeidx}  
\usepackage{url}
\usepackage[square, numbers, comma, sort&compress]{natbib} 

\usepackage{float}
\usepackage{algorithm}
\usepackage{multirow}

\usepackage{amsthm}
\usepackage{amsmath}
\usepackage{graphicx}
\usepackage{array}
\usepackage{lscape}
\usepackage{caption}
\usepackage{verbatim}
\usepackage{textcomp}
\usepackage{mathtools}

\floatname{algorithm}{Protocol}

\newtheorem{definition}{Definition} 
 
\newtheorem{theorem}{Theorem} 
\newtheorem{lemma}{Lemma}

\def\ket#1{| #1 \rangle}

\def\bra#1{\langle #1 |}

\definecolor{darkgreen}{RGB}{10,150,50}

\begin{document}
\allowdisplaybreaks[3]
\frenchspacing

\title{\bf{On optimising quantum communication \\ in verifiable quantum computing}}

\author[1]{Theodoros Kapourniotis}
\author[1,2,3]{Vedran Dunjko}
\author[1,4]{Elham Kashefi}
\affil[1]{School of Informatics, University of Edinburgh, UK}
\affil[2]{Division of Molecular Biology, Ruder Boskovic Institute, Zagreb, Croatia}
\affil[3]{Now at: Institute for Theoretical Physics, University of Innsbruck, Austria}
\affil[4]{Now at: CNRS, TelecomParis Tech, France}

\maketitle
\begin{abstract}

In the absence of any efficient classical schemes for verifying a universal quantum computer, the importance of limiting the required quantum resources for this task has been highlighted recently. Currently, most of efficient quantum verification protocols are based on cryptographic techniques where an almost classical verifier executes her desired encrypted quantum computation remotely on an untrusted quantum prover. In this work we present a new protocol for quantum verification by incorporating existing techniques in a non-standard composition to reduce the required quantum communications between the verifier and the prover.

\end{abstract}

\section{Introduction}

As quantum technology scales, verifying that a quantum black box device does what it promises becomes a difficult problem \cite{falsifiable}. In particular, verification implies that the probability of accepting the output returned by a remote quantum computation and the output being incorrect, can be made arbitrarily small. Providing a solution to this problem is crucial in the context of secure delegated quantum computing where a weak verifier is required to perform a complex computation with the help of an untrusted quantum prover \cite{Childs,arrighi06traps,broadbent2009universal,fitzsimons2012unconditionally,aharonov2008interactive,reichardt13classical,broadbent13onetime,morimae13measurement}. Also, in lab settings, where an experimentalist wants to verify that the state returned by a scalable quantum experiment is correct, standard techniques like quantum tomography become intractable, for few examples see \cite{bosonsampling,aolita2014reliable,spagnolo2014experimental,GAN14}.

To achieve a practical verification protocol (from the verifier's point of view) one aims to minimise the required resources of the verifier. This means optimising the number of operations that the verifier has to apply as well as reducing the classical and quantum communication rounds between verifier and prover, while limiting the quantum capabilities of the verifier in terms of the states preparation and quantum operations he can apply. Recently several verification protocols have been proposed, each targeting particular aspects of the above optimisation goal \cite{arrighi06traps,barz2013experimental,aharonov2008interactive,fitzsimons2012unconditionally,reichardt13classical,broadbent13onetime,mckague,Hajdu,gheorghiu15hybrid1,hayashi2015verifiable}. 

In this paper we are concerned with two such protocols, one with an optimal state preparation, proposed by Fitzsimons and Kashefi (referred to here as the FK protocol \cite {fitzsimons2012unconditionally}), the other with better communication cost scaling proposed by Aharonov, Ben-Or and Eban (referred to here as the ABE protocol and is the second of the two protocols presented in \cite{aharonov2008interactive}). In the FK protocol the verifier is required to have the minimal quantum capacity of preparing single random qubits, while due to the complexity of the underlying resource state necessary for the prover's universal quantum computation, the communication overhead is quadratic in the input size. In contrast, the ABE protocol enjoys a linear overhead in the input size however the cost is the verifier's requirement of preparing highly entangled secretly encoded states which scale with the security parameter. The main contribution of this paper is to show how we can keep the best of the two approaches. In doing so we ought to prove several new properties of these two protocols that could become useful in other context as well. 

Our FK-ABE hybrid protocol is based on the blindness property of the FK protocol where the prover does not learn anything about the computation. Hence, the verifier can remotely and securely prepare the entangled encoded state needed for the ABE protocol on the prover side. After this preparation stage is performed via the FK protocol, without obtaining the state back, verifier continues with the steps of the ABE protocol to perform the (public) logical computation and check if the output state is valid. Crucially, as we prove, the techniques of the FK protocol can be used separately for the preparation of each encoded input state of the ABE protocol, avoiding the construction of a global quadratic size resource state used in the original FK protocol. Hence the overall complexity of the hybrid protocol remains linear in the input size while we get the advantage of preparing single qubit states (in this case qudit to match the ABE protocol) as opposed to the highly entangled encoded states required for the original ABE protocol. However, several issues had to be resolved in order to construct the hybrid protocol.  

\begin{itemize}
\item The verification procedure of the FK protocol involves measuring special \emph{trap} qubits that are placed at random positions including among the output system (the final qubits not measured by prover). In order to keep the verifier as simple as possible, i.e. not requiring any quantum memory capability, unlike the original FK protocol, the final output state prepared at the first stage of the hybrid protocol is not returned to the verifier. However the position of the actual output of the computation should be made public so that prover can apply the public logical circuit of the ABE protocol. This is achieved by adding extra blind computation steps that localise the position of the output without leaking the positions of the traps.

\item The final output of the modified localising protocol, as explained above, by construction does not contain any traps. Therefore the prover can replace this state by any state of his choice and there is no mechanism for the verifier to detect his deviation. However, we show that, due to blindness, the output state has the correct structure to be verifiable by the ABE protocol stage of the hybrid protocol and can possibly be used in any other hybrid compositions that involve similar verification techniques.

\item The ABE protocol has to be proved to be robust, since due to the above mentioned issues, the output prepared by the FK protocol stage could be deviated from the ideal ABE protocol input. We show that as long as the deviation is independent of any secret parameter of the verifier the ABE protocol succeeds.

\item Finally, we also prove that the classical communication between verifier and prover, necessary for executing the logical circuit computation in the ABE protocol, does not leak any information that violates the blindness requirement in the context of the hybrid protocol, thus not affecting the verification property.
\end{itemize}

\subsection{Related work} 

A different verification approach was proposed in \cite{reichardt13classical} where the verifier is fully classical but the computation is delegated to non-signalling entangled provers and the communication overhead is huge due to the rigidity requirement for testing CHSH inequalities. While the communication of this protocol has been improved recently \cite{gheorghiu15hybrid1,Hajdu}, also through a hybrid composition with the FK protocol, however the bound achieved there is substantially higher (at least quartic) than the linear bound of the presented protocol. Other work on quantum verification includes \cite{arrighi06traps} where there is a restriction on the permitted set of functions, referred to as random verifiable. In \cite{broadbent13onetime} a verification protocol is presented with the focus being on constructing a quantum one time program rather than improving on the practicality of verification. In \cite{hayashi2015verifiable} a verification protocol is given where the prover is able to send quantum states back to the verifier which has the capacity to measure them. Focusing on the blindness property only, several recent papers \cite{GMMR13,MPF13,PF14} have achieved better than linear bound while compromising either unconditional security property or simplicity of the verifier quantum operation. We believe that it may be worthwhile to explore these new blindness techniques with the goal of designing more efficient verification protocols.

Before formalizing the main results of this paper, we give a short overview of the FK and ABE protocols.

\subsection{Overview of the FK and ABE protocols} \label{1.1}

A security property that is defined in the delegated computing scenario and used as a stepping stone in the construction of the FK protocol is blindness \cite{fitzsimons2012unconditionally}. The verifier (Alice) wants to delegate a computation to the prover (Bob) while hiding both the input and the computation. Bob's possible deviation is not constrained in any way.

\hskip 0.1cm
\begin{definition} [Perfect Blindness] \label{def:blindness} Let $P$ be a protocol for delegated computation: Alice's input is a description of a computation on a quantum input, which she needs to perform with the aid of Bob and return the correct quantum output. Let $\rho_{AB}$ denote the joint initial state of Alice and Bob and   $\sigma_{AB}$ their joint state after the execution of the protocol, when Bob is allowed to do any deviation from the correct operation during the execution of $P$, averaged over all possible choices of random parameters by Alice. The protocol $P$ is perfectly blind if
\begin{gather}
\forall \rho_{AB} \in \mathcal{L} (\mathcal{H}_{AB}), \exists \mathcal{E}:\mathcal{L}(\mathcal{H}_B) \rightarrow \mathcal{L}(\mathcal{H}_B)  \text{, s.t. } \;\;\; \text{Tr}_A (\sigma_{AB}) = \mathcal{E} (\text{Tr}_A(\rho_{AB})) \label{eq11}
\end{gather}
\end{definition}

In the verification cryptographic setting Alice wants to delegate a quantum computation to Bob and accept or reject the result depending on whether the returned outcome is correct, or Bob has deviated \footnote{In this work we will assume that the apparatuses of Alice and Bob are perfect. Without this assumption on Bob, the protocol will also detect any errors which may stem from Bob's faulty devices.}. This definition only concerns pure state inputs and unitary computations.

\hskip 0.1cm
\begin{definition}[Verifiability]  \label{def:verifiability}
A protocol for delegated computation is $\epsilon$-verifiable ($0\leq \epsilon < 1$) if for any choice of Bob's strategy $j$, it holds that for any input of Alice:
\begin{equation}
\text{Tr} (\sum_{\nu} p(\nu) P^{\nu} _{\text{incorrect}} B_j (\nu)) \leq \epsilon
\end{equation}
where $B_j(\nu)$ is the state of Alice's system A at the end of the run of the protocol, for choice of Alice's random parameters $\nu$ and Bob's strategy $j$. If Bob is honest we denote this state by $B_0 (\nu)$. Let $P_{\perp}$ be the projection onto the orthogonal complement of the the correct quantum output. Then,
\begin{equation}
P^{\nu} _{\text{incorrect}} = P_{\perp}\otimes \ket{ACC}\bra{ACC}
\end{equation}
where $ \ket{ACC}$ is the accept state for the indicator that Alice sets at the end of the protocol.
\end{definition}

The main idea of the FK protocol is to test prover's honesty by hiding trap qubits among all qubits sent by verifier, in a way that they do not affect the result of the actual computation. Blindness means that prover cannot learn the position of the trap, nor its state. During the execution of the measurement based computation prover is asked to measure this trap qubit and report the result to verifier. If prover is honest this measurement gives a deterministic result, which can be verified by verifier and hence a dishonest sever will be detected with no-zero probability. Moreover the computation is encoded in a quantum error correcting code (QECC) correcting up to $d$ errors, forcing malicious Bob to cause larger errors, if he is to interfere with the computation. These larger errors, in turn, are more easily detected.

\hskip 0.1cm 
\textbf{Sketch of the FK protocol} (Protocol 8 in \cite{fitzsimons2012unconditionally})

\begin{enumerate}
\item Alice has a description of a computation in the measurement-based quantum computing (MBQC) model. She encodes the computation in a fault tolerant MBQC pattern, using a QECC that detects $d$ errors.
\item Alice designs classically the embedding procedure of the encoded computation into a suitable MBQC pattern where the vertices can be divided into subsets $S_{\gamma}$ where each subset contains a computational qubit and a trap qubit placed at a random position inside the subset. For the purpose of separating the computational and the trap qubits, special `dummy' qubits prepared in the computational basis are injected.
\item Alice prepares all single qubits needed to construct the underlying entangled state for the pattern computation, pre-rotated by a random angle around $Z$ axis in order to hide the measurement angle.
\item Bob receives the qubits one by one and entangles them to construct the  underlying entangled state.
\item Alice sends to Bob at each step the measurement angles that implement the computation adapted based on the previous measurements results as well as cancelling the pre-rotation while adding a bit flip on the outcome to hide the measurement result.
\item Alice receives the measurement result and corrects it.  
\item The isolated traps qubits are measured in the basis of their pre-rotated angles so the outcome is deterministic if Bob is honest.
\item At the end Alice accepts the outcome returned by Bob if all traps give the correct outcome.
\end{enumerate}

\hskip 0.1cm

In the following Lemma a distinction is made between the off-line phase of the protocol which Alice and Bob can execute before Alice receives the actual input, which is the description of the computation, and the on-line phase in which Alice and Bob repeatedly communicate to perform the measurements and corrections on the entangled state.

\hskip 0.1cm

\begin{lemma} The FK protocol is perfectly blind (Theorem 12 in \cite{fitzsimons2012unconditionally}) and $\epsilon$-verifiable where $\epsilon=\frac{1}{c^{d_1}}$  (Theorem 13 in \cite{fitzsimons2012unconditionally}) where $d_1$ is the number of errors detectable by the QECC used for amplification and some constant $c>1$. The quantum requirement of Alice is to prepare single qubit states and the communication requirement is $\widetilde{O}(n^2) \times O(log(1/\epsilon))$ separable single qubit states sent from Alice to Bob off-line and $\widetilde{O}(n^2) \times O(log(1/\epsilon))$ bits of on-line classical communication between Alice to Bob,  where $n$ is the size of the computation.
\end{lemma}

The number of of errors $d_1$ detectable by the QECC used for amplification will be referred to as the security parameter of the FK protocol in the rest of this paper.

In the ABE protocol the input is encoded by Alice using a secretly randomised QECC with specific properties that allow for the secret updating of the random key throughout the execution of the computation. It is then hard for Bob, without the knowledge of the key to deviate while keeping the state inside the legal subspace. The computation performed on the encoded input can be publicly announced without compromising the verifiability of the protocol. 

\hskip 0.1cm
\textbf{Sketch of the ABE protocol} (Protocol 4.2 in \cite{aharonov2008interactive})
\begin{enumerate}
\item Alice has a description of a computation in the gate teleportation model  where the non-Clifford operations are performed using non-Clifford magic states, Clifford gates and computational basis measurements.
\item  She encodes the input (typically the blank state) to the computation in a qudit polynomial-QECC code and translates the computation to a logical one where a universal set of gates can be implemented transversally in this computation model. We refer to this public encoding of the computation as the public polynomial-QECC logical circuit.
\item She applies two layers of randomization on each logical input qudit: a random sign on each of the physical qudits, the same sign key is reused for all the different logical qudits, and a random Pauli rotation (different for each logical qudit).
\item Bob performs the computation while classically communicating with Alice to implement the corrections needed after each measurement step.
\item Alice receives the final output, undoes the quantum one time pad and the random sign key, using updated secret keys depending on the public polynomial-QECC logical circuit and the communication with Bob,  and checks if it is inside the valid subspace of the code. If not, she rejects.
\end{enumerate}

\hskip 0.1cm
\begin{lemma} The ABE protocol is $\epsilon$-verifiable where $\epsilon=\frac{1}{2^{d_2}}$ (based on Theorem 4.1 in \cite{aharonov2008interactive} and ignoring the circuit error probability) where $d_2$ is the degree of polynomials in the polynomial-QECC. The quantum requirement of Alice is to prepare $O(n)$ states of $O(1/log(\epsilon))$ entangled qudits of $O(1/log(\epsilon))$-level  and the communication requirement is $O(n)$ states of $ O(log(1/\epsilon))$ entangled qudits sent from Alice to Bob off-line and $O(n) \times O(log(1/\epsilon))$ bits of on-line classical communication between Alice to Bob, where $n$ is the size of the computation.
\end{lemma}

The degree $d_2$ of polynomials used in the polynomial-QECC will be referred to as the security parameter of the ABE protocol in the rest of this paper.

\section{Main results}

We propose a new verification protocol that combines the techniques of the FK and ABE protocols. The main idea is to use the verifiable computation of the FK protocol to prepare on the prover's side the inputs encoded by the randomised polynomial QECC, that is, the states Alice would have to prepare and send in the first step of the original ABE protocol.  On these states the public logical circuit of the ABE protocol is applied and the output is tested for being in the correct encoded subspace.  If either of the trapification or decoding procedures fails, the verifier rejects. The partitioning of the underlying resource state used in the FK protocol into smaller sub-states for each separable logical qudit leads to an improved communication complexity of the hybrid protocol compared to the original FK protocol.

In Section 2.1 we give an overview of how the FK protocol is modified for the needs of the preparation of the encoded state of the ABE protocol. Section 2.2 examines the problems of incorporating the computation and detection procedures of the ABE protocol into the hybrid protocol. In Section 2.3 we overview the hybrid protocol and how it achieves its improved complexity. The full description of the results is given in later sections and the detailed technical proofs are all in the Appendix.

\subsection{Modifying the FK protocol for remote state preparation}

We use the main detection and amplification technique of the FK protocol to remotely prepare the state needed for the ABE protocol on Bob's system. Due to blindness, Bob does not learn anything about this state and therefore any secret random parameters it contains, including a quantum one-time-pad with a key selected by Alice.  The requirement for this state to be useful for the detection procedure of the ABE protocol is that it can be written in terms of the correct encoded state up to an arbitrary CPTP-map which does not depend on the secret parameters. We call such states verifiable states.

Two problems had to be tackled in modifying the FK protocol for this purpose.  The first problem is that in the FK protocol, it is impossible to help the server distinguish the computational output systems from traps and dummies without jeopardizing the FK protocol's verifiability. We therefore construct a simple protocol to blindly transform the output of the FK protocol to the input that contains only the computational output system at a fixed position. Hence, Bob can apply the ABE protocol logical circuit on the output state while not learning anything about the trap qubits. Secondly, since the state at the end of the preparation phase does not contain any traps there is no possibility to test for the correctness of the output. However, since we only require  a relaxed variant  of verifiability (correctness up to a fixed CPTP map)  this is proved not to be a problem by modifying the original proof of verification. The modified protocol with the localisation gadget we call the localising protocol.

\hskip 0.1cm 
\textbf{Sketch of the localising protocol}\footnote{Full protocol is given in Section 4}

\begin{enumerate}
\item Alice and Bob follow the same steps as in the FK protocol, but an extra `gadget' state (depicted in Figure \ref{gadget}) is attached on the output system qubits of the entangled state prepared by Bob for the original FK protocol. This gadget state does not contain any dummy or trap qubits on its final layer therefore effectively fixing the position of the output.
\item After all qubits of the normal entangled state are measured and the traps are tested Bob is asked to measure the remaining qubits in a way that only the output qubits are teleported to the last layer.
\item The state that Bob holds at the end is quantum one-time-padded with a key known only to Alice.
\end{enumerate}

\hskip 0.1cm 
\begin{theorem} \label{theorem1} The output of the localising protocol after decoding of the QECC  utilized in the FK protocol and decoding of the quantum one-time-pad, averaged over all random parameters of Alice, denoted as $\nu$, is of the following form:

\begin{equation*}
\sum_{\nu} \rho_{out}^{\nu} \approx_{\epsilon} p_{acc} \mathcal{E} ( \ket{\psi_{c}}\bra{\psi_{c}} ) \otimes \ket{ACC} \bra{ACC} + (1-p_{acc}) \rho' \otimes \ket{REJ}\bra{REJ}
\end{equation*}

where $\epsilon =  \frac{1}{c^{d}}$ for some $c>1$ and $d$ the number of errors corrected by the FK protocol QECC, $p_{acc}$ is a probability, $\ket{\psi_{c}}$ is the correct  output of the computation, $\mathcal{E}$ is a CPTP-map where all Kraus operators are Pauli operators and is independent of $\ket{\psi_{c}}$ and $\rho'$ is a normalized state.
\end{theorem}

It is important to note that using only a blind quantum computing protocol, without inserting any trap qubits for verification, one could not obtain the above required correct form for the final output system. In particular the possible cumulative perturbations that Bob may cause may be correlated to the state they act on, hence may not be expressible as a fixed CPTP map.

\subsection{Incorporating the verification procedure of the ABE protocol}

As mentioned before, in the hybrid protocol, Bob does not return the state to Alice after the localising protocol but he applies the polynomial-code logical circuit to implement the actual computation on the encoded state. In order to incorporate these extra steps we have to resolve a number of problems.

Firstly, the state that Bob holds is encoded by the QECC used for the amplification of the detection probability in the FK protocol. In order to apply the logical circuit of the ABE protocol  on this state we need to ask Bob to decode the first QECC. Fortunately the QECC of the FK protocol is based on a stabiliser code with a Clifford decoding circuit \cite{raussendorf2007fault}. The decoding circuit is publicly known and there is no need for any communication between Alice and Bob. An honest Bob performs the operators and Alice simply updates her quantum one-time-padding secret keys accordingly. Hence no information is leaked on the secret keys, also uniformity of the keys is preserved.

Secondly, the ABE protocol requires quantum states of prime dimension $q$ where $q>2$, whereas the original FK protocol uses qubits. To resolve this we present an analogue of the FK protocol which can use systems of any dimension while preserving all the security properties. The resources requirement for Alice is the same, the ability of preparing random single qudit states and the overall communication complexity remains the same as the original FK protocol. 

Thirdly, since the computation we use from the ABE protocol is based on gate teleportation, there is a constant number of rounds of classical communication between Alice and Bob per non-Clifford gate teleportation. We prove, however, that this communication does not leak any of the secret parameters of the trapification phase, crucial for the security of the hybrid protocol.

Finally, we prove since the detection and decoding procedure that Alice applies to the returned state of the ABE protocol is a CPTP map, the resulting state of the overall protocol is $\epsilon$-close to the final state of the ABE protocol, thus also verifiable.

Therefore we have the necessary tools to build a hybrid protocol by incorporating in the preparation stage (with the localising gadget) the application of the decoding part so that the assumptions of Theorem \ref{theorem2} can be modified to arrive to the desired verification property. However this protocol would not yet give any improvement in terms of complexity over the existing protocols and to achieve this an extra step has to be employed as we describe next.

\subsection{Hybrid Protocol}

We partition the underlying entangled state necessary for the FK protocol phase of the hybrid protocol into smaller separable sub-states preparing each a polynomial-QECC encoded input state of the ABE protocol. Since the quadratic round complexity of the FK protocol comes from the complex structure of the overall required underlying resource state, simplifying the state for the purposes of this hybrid composition (observing that the polynomial-QECC encoded logical inputs are separable) succeeds in reducing the communication complexity. This modification does not violate the verification properties of the FK protocol since the fact that the states are separate is public knowledge and the detection and encoding procedures for each of the separate states can be done locally. The sketch of the hybrid protocol is given below:

\hskip 0.1cm 
\textbf{Sketch of the FK-ABE hybrid protocol}\footnote{Full protocol is given in Section 5}
\begin{enumerate}
\item Alice and Bob execute a specified number of instances of the qudit analogue of the FK protocol (that includes the localising gadget) to prepare in Bob's system each of the desired polynomial-QECC state of the ABE protocol.
\item Bob applies the publicly known Clifford decoding circuit for the QECC used for the amplification of the error detection probability of the FK protocol, and Alice updates her one-time-pad keys. 
\item Alice and Bob apply the polynomial-QECC logical circuit on the encoded states following the steps of the ABE protocol. Alice updates her one-time-pad keys.
\item Bob sends the final measurement outcomes to Alice and Alice employs the detection and decoding procedure of the randomised polynomial-QECC (which is the final step of the ABE protocol as well).
\end{enumerate}

\hskip 0.1cm
\begin{theorem} \label{theorem2}
Hybrid protocol is correct and $\epsilon$-verifiable, with $\epsilon \leq \epsilon_1 + \epsilon_2$ where  $\epsilon_1 = \frac{1}{c^{d_1}}$, $\epsilon_2 = \frac{1}{2^{d_2}}$ where $d_1$ and $d_2$ are the security parameters of the FK and ABE protocol phase of the hybrid protocol respectively and $c>1$ is a parameter of the FK protocol phase. Alice needs only to prepare random single qudit states and the required classical and quantum communications is of order $O(n \text{Polylog}(\frac{1}{\epsilon}))$, where $n$ is the size of the computation.
\end{theorem}

The steps which are essential for the construction of this protocol  can potentially be used for constructing different hybrid schemes.

\section{Preliminaries}

The FK protocol is defined in the measurement based quantum computing model (MBQC), basic elements of which are outlined in Section \ref{prem1}. The ABE protocol is a $d$-level quantum system protocol with $d>2$ and is based on the Gate Teleportation Model outlined in Section \ref{prem2}. Finally the Quantum Error Correcting Code used in the ABE protocol is given in Section \ref{prem3}.

\subsection{Measurement Based Quantum Computing} \label{prem1}

We introduce the notation necessary to describe a computation in MBQC \cite{RB01,danos2007calculus}. A generic computation consists of a sequence of commands acting on qubits:
\begin{itemize}
\item $N_i(\ket{q})$: Prepare the single auxiliary qubit~$i$ in the state $\ket{q}$;

\item $E_{i,j}$: Apply entangling operator $cZ$ to qubits $i$ and $j$;

\item $M^{\alpha}_i$: Measure qubit $i$ in the basis $\{ \frac{1}{\sqrt{2}} ( \ket{0} + e^{i \alpha} \ket{1} ), \frac{1}{\sqrt{2}} ( \ket{0} - e^{i \alpha} \ket{1} ) \}$ followed by trace out the measured qubit. The result of measurement of qubit $i$ is called signal and is denoted by $s_i$;

\item $X_i^{s_j}, Z_i^{s_j}$: Apply a Pauli $X$ or $Z$ correction on qubit $i$ depending on the result $s_j$ of the measurement on the $j$-th qubit.

\end{itemize}
The corrections could be combined with measurements to perform `adaptive measurements' denoted as $\prescript{s_z}{}{\left[ M^{\alpha}_i \right]}^{s_x} = M^{(-1)^{s_x}\alpha+s_z \pi}_i $. A computation is formally defined by the choice of a finite set $V$ of qubits, two not necessarily disjoint sets $I \subset V$ and $O \subset V$ determining the inputs and outputs, and a finite sequence of commands acting on $V$.

The entangling commands $E_{i,j}$ define an undirected graph over $V$ referred to as $(G,I,O)$. Along with the pattern we define a partial order of measurements and a dependency function $D$ which is a partial function from $O^C$ to $\mathcal{P}^{I^C}$, where $\mathcal{P}$ denotes the power set. Then, $j \in D^x_i$ if $j$ gets a Pauli $X$ correction depending on the measurement outcome of $i$ and $j \in D^z_i$ if $j$ gets a Pauli $Z$ correction depending on the measurement outcome of $i$. A sufficient condition on the geometry of the graph state to allow unitary computation is the existence of flow (\cite{danos2006determinism}). In what follows, $x \sim y$ denotes that  $x$ is adjacent to $y$ in $G$.

\hskip0.1cm
\begin{definition} \label{d-flow} \cite{danos2006determinism} A \emph{flow} $(f,\preceq)$ for a geometry $(G,I,O)$ consists of a map $f:O^c\mapsto I^c$ and a partial order $\preceq$ over $V$ such that for all $x\in O^c$
\begin{itemize}
\item[(F0)]~~$x \sim f(x)$;
\item[(F1)]~~$x \preceq f(x)$;
\item[(F2)]~~for all $y\not = x, y \sim f(x)$:  $x \preceq y$\,.
\end{itemize}
\end{definition}

\subsection{$d$-level quantum operations and gate teleportation model}\label{prem2}

First we describe the d-level generalized gates that will be useful in the following constructions. All additions and multiplications in these definitions are performed modulo $d$, unless otherwise stated. The generalized Pauli operators over $\mathbf{F}_d$ are defined as: $X\ket{a}\equiv\ket{a+1}$, $Z\ket{a}\equiv\omega_{d}^{a}\ket{a}$, where $\omega_{d}=e^{2\pi i/d}$ is the primitive $d$th-root of unity. The generalized Hadamard (or Fourier) gate F is defined as:

 \begin{displaymath} \ket{a} \mapsto \frac{1}{\sqrt{d}} \sum_{b \in \mathbf{F}_d} \omega _{d}^{ab} \ket{b}. \end{displaymath}
 
The $cX$ is defined as: $\ket{a,b} \mapsto \ket{a,a + b}$ and $cZ$ as: $\ket{a,b} \mapsto \omega_d^{ab} \ket{a, b}$.
The generalized phase gate S is defined as:

\begin{displaymath} \ket{a} \mapsto \omega_d^{\frac{(a+1)a}{2}}\ket{a} \end{displaymath}

The generalized Toffoli gate which complements the set of Clifford gates to provide universality (\citep{howard2012qudit}, \cite{anwar2014fault}) is defined as: $\ket{a,b,c} \mapsto \ket{a,b,c + ab}$. An alternative to the Toffoli gate to complement Clifford gates for universality is the family of `so-called' generalized `$\pi /8$' gates \citep{howard2012qudit} one of which, used in this paper, is defined as: $T\ket{a}\equiv\omega_{d}^{a^3}\ket{a}$ with the exception of qutrits where $T_3\ket{a}\equiv\omega_{9}^{(a^3 \text{ mod } 9)}\ket{a}$.

In the gate teleportation model the ABE protocol is defined in, all Clifford gates are performed by applying unitary operators, while the non-Clifford gates are performed by entangling with magic non-Clifford states (Toffoli states in the case of the ABE protocol), measuring in the computational basis and performing Clifford correction operators on the remaining states \cite{nielsen2000book}.

\subsection{Polynomial Quantum Error Correcting Code}\label{prem3}

The class of quantum error correcting codes used in the ABE protocol, is based on polynomial error-correcting codes \cite{aharonov1997code}. By randomizing these codes using a secret sign key one can acquire a Quantum Authentication Scheme (QAS) \cite{barnum2002qas}, usable by two parties who want to authenticate a quantum state exchanged through a malicious quantum channel. Due to specific properties of the \emph{signed} polynomial error-correcting codes, they can also be used  in the context of secure quantum multiparty computation \cite{ben2006secure} and in the context verification of a single quantum prover as demonstrated in the ABE protocol.

Specifically, the code operates over the field $\mathbf{F}_d$, where $d$ is prime and represents the dimension of the quantum system. The code uses polynomials $f(\cdot)$ with coefficients from $\mathbf{F}_d$ and with a maximum degree of $p$, where $p$ is a parameter of the code. These polynomials are evaluated at $m$ \emph{distinct non-zero points} from $\mathbf{F}_d$: $\{a_1 \ldots a_m\}$. To be able to choose $m$ distinct non-zero points from $\mathbf{F}_d$ we impose the restriction: $m<d$. Moreover, we set $m=2p+1$. Finally, the code uses a secret sign key that is a uniformly randomly chosen string ${\bf k} \leftarrow \{\pm 1 \}^m$.

Given a quantum state $\ket{a}$, where $a \in \mathbf{F}_d$, we can encode it using the signed polynomial error-correcting code:

\begin{displaymath} \ket{a} \rightarrow \frac{1}{\sqrt{q^p}} \sum _{\{f|deg(f)\leq p,f(0)=a\}} \ket{(k_1 f(a_1), \ldots , k_m f(a_m))\mod{d}} \end{displaymath}

where the mod $d$ is taken element-wise in the label of the right-hand side ket.

To detect if there was any Pauli error in the transmitted codeword, the receiver undoes the sign key and checks if the resulting string is the representation, at $m$ distinct points, of a polynomial of maximum degree $p$. If not it aborts and declares the state invalid. The probability of accepting an incorrect state (when the channel applies an arbitrary Pauli attack) is exponentially small on $p$. In order to acquire the same property for any type of malicious attack, an extra quantum one-time-pad is added on top of the existing randomization, as it is the case in the ABE protocol.

\begin{algorithm}

\caption{ Localising Protocol (based on the FK protocol)  \label{prot:vubqc_notraps}}

\hskip0.2cm

\textbf{Alice's input.} Description of a computation in the MBQC model (or equivalent) using a convenient underlying open graph state $(G,I,O)$. The computation is represented, for any qubit $i \in G \setminus O$, as a measurement angle $\varphi_i$ (together with the set of X-dependences $D^X_i$ and Z-dependences $D^Z_i$ and a fixed partial order of measuring depending on the graph structure). The input is set to be the Hadamard basis state of $n$ qubits: $\ket{+}^{\otimes n}$. Protocol can be extended to admit external quantum input by applying the techniques described in the FK protocol.

\hskip0.2cm

\textbf{Bob's output.} A quantum state that contains the encoded and one time padded quantum output of the computation.

\hskip0.2cm

\textbf{The protocol}

\begin{enumerate}
\item Preprocessing 1. Alice translates the computation to a Fault Tolerant (FT) MBQC pattern that can detect $d$ errors. Let the updated open graph be $(G',I',O')$, where $|G|=m'$ and $|I'|=|O'|=n'$. \label{step1}
\item Preprocessing 2. Alice embeds the encoded computation pattern into a suitable graph which has the following property: There exists a fixed order of measurement which respects the computational flow and each computational qubit belongs to a constant size subset of qubits $S_{\gamma}$ in which a trap can be at any position with uniformly random probability. An example is the dotted complete graph of size $O(m'^2)$. \label{step2}
\item Preprocessing 3. Alice attaches a gadget graph (Figure \ref{gadget}) on each subset $S_{\gamma}$ of qubits of the output system of the previous step's graph. The effect of these extra graphs is that they fix the position of system $O'$. The total number of qubits of the final graph is $N$. \label{step3}
\item Alice prepares the rotated qubits. For $i=1$ to $N$: \label{step4}

\begin{enumerate}
\item If qubit is a dummy: prepares $\ket{d_i}$, $d_i \leftarrow_R \{0,1\}$ (where $\leftarrow_R$ means \emph{chosen uniformly at random from}). \label{step4a}

\item If qubit is not dummy and not in $O'$: prepares  $\prod_{j \in N_G(i)\cap D} Z^{d_j} \ket{+_{\theta _{i}}}$, where $\ket{+_{\theta _{i}}}\equiv \frac{1}{\sqrt{2}}(\ket{0}+e^{i\theta _{i}} \ket{1})$,  $\theta _{i} \leftarrow_R \{0, \pi /4, 2\pi /4, \ldots, 7\pi /4\}$. \label{step4b}

\item If qubit is in $O'$: Same as previous step but $\theta$ is fixed to 0 and an extra $Z^{r_i}$ rotation is applied where $r_i$ is a random bit. \label{step4c}

\item She sends the qubit to Bob. \label{step4d}

\end{enumerate}

\item Bob entangles the states according to the graph state by applying $cZ$ gates. \label{step5}

\item Bob performs the rest of the computation using classical help from Alice. For $i$ which ranges over all qubits (respecting the flow) except system $O'$:

\begin{enumerate}

\item Alice computes the actual measurement angle $\phi ' _{i}$ using the dependences and the previous measurement results ($\phi_i = 0$ for dummies and traps). \label{step6a}

\item Alice chooses a random bit $r_{i}$ and computes $\delta _{i}=\phi ' _{i} + \theta _{i} +\pi r_{i}$. \label{step6b}

\item Alice transmits $\delta _{i}$ to Bob. \label{step6c}

\item Bob performs measurement $M_{i}^{\delta _{i}}$ on qubit $i$. \label{step6d}

\item Bob transmits the result to Alice. \label{step6e}

\item Alice flips the result if $r_{i}=1$, otherwise does nothing. \label{step6f}

\end{enumerate}

\item Alice sets her indicator bit to accept if all trap tests where successful. \label{step7}

\item Bob's system will contain the output qubits $O'$ placed at a fixed position on the graph.  \label{step8}

\end{enumerate}

\end{algorithm}

\section{Localising Protocol} \label{section_step}

In this section we give an explicit description of the localising protocol. The core methodology is the same as in the FK protocol. Each trap is prepared in state $\ket{+_{\theta_t}}$, where $\theta$ is chosen at random from angle set $A=\{0, \pi /4, 2\pi /4, \ldots, 7\pi /4\}$, and placed at position $t$ in the graph. Positions $t$ are such that for the graph used in the FK protocol there exists a partitioning into sub-graphs $S_{\gamma}$ that the each trap is placed at a uniform random position inside each $S_{\gamma}$. In the FK protocol the graph that is used to achieve this property is the dotted-complete graph \cite{fitzsimons2012unconditionally}, while other graphs can possibly satisfy the same property. In order to separate the qubits participating in the computation and the trap qubits, dummy qubits prepared in the computational basis are placed between them so that the corresponding entangling operators have no effect. The choice of the basis for each dummy state is uniformly random to satisfy the blindness property. In particular, if a dummy qubit is in state $\ket{0}$, applying the entangling operator $cZ$ between this qubit and a qubit prepared on the $(X,Y)$ plane has no effect. If a dummy qubit is in state $\ket{1}$  then applying $cZ$ will introduce a Pauli $Z$ rotation on the qubit prepared on the $(X,Y)$ plane. This effect can be cancelled by Alice in advance, by introducing a Pauli $Z$ rotation on all the neighbours of $\ket{1}$'s when preparing the initial state. 

During the execution of measurement based computation Bob is asked to measure the trap qubits, at positions $t \in T$, each with angle $\theta_t+r \pi$ and return each classical result $b_t$ to Alice. If for all trap qubits $t \in T$ we have $b_t=r_t$ Alice sets an indicator bit to state $acc$ (which means that this computation is accepted), otherwise she sets it to $rej$ (computation is rejected). Due to blindness (achieved by the random rotations) the prover cannot learn the position of traps so he is not able to return with probability one the correct results of the trap measurements while deviating from the steps of the protocol.  The computation is encoded in a fault tolerant procedure using a QECC that can perfectly correct $d$ errors, so dishonest prover has to attack at more than $d$ positions, thus having exponentially small in $d$ probability of succeeding in predicting the correct results of the corresponding traps.

\begin{figure}[h!] 
  \centering
      \includegraphics[width=0.5\textwidth]{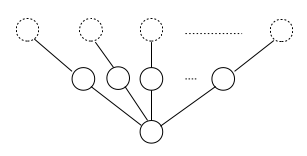}
  \caption{Single gadget state attached to a subset $S_{\gamma}$ (represented by dashed circles) of the output of the normal graph used in the FK protocol. The number of branches depends on $|S_{\gamma}|$. Only one of the qubits of the first row of the gadget is a non-dummy state allowing to teleport blindly the actual output to the bottom qubit.    \label{gadget}}
\end{figure}

\hskip 0.1cm

The main difference from the original FK protocol is the modification of the graph state by appending on the output a specific number of gadget states, one of which is depicted in Figure \ref{gadget}.  Specifically one gadget is attached on each subset $S_{\gamma}$ of the output of the FK protocol graph (system to be normally returned to Alice). Following the construction in the FK protocol, we assume that each $S_{\gamma}$ contains one trap qubit, a number of dummy qubits and one qubit participating in the actual computation (in the dotted-complete graph based version of the FK protocol $\forall \gamma,|S_{\gamma}|=3$). The goal of the gadget system is to teleport only the computational qubit to the fixed vertex at the bottom of each gadget. Therefore the first layer of the gadget qubits (in Figure \ref{gadget} the first row of solid circles, that are connected directly to the existing graph of the FK protocol)  contains $|S_{\gamma}|-1$ dummy qubits which isolate  the trap and the dummies of the output $S_{\gamma}$ from the bottom qubit. The computational qubit and the gadget qubit that is connected to it are both measured with $\phi=0$ so that the effective computation is identity and the qubit is teleported to the bottom of the gadget. The gadget structure is such (one gadget per $S_{\gamma}$ and symmetric) that they can be added without breaking the uniformity of the positioning of the traps in the FK protocol graph. The steps of the localising protocol are given in Protocol \ref{prot:vubqc_notraps}.

\hskip 0.1cm

A sketch of the proof  of  verifiable state preparation, formally described in  \ref{theorem1}, is given here, while a detailed proof is presented in Appendix \ref{app_a}.

The following lemma will be used in the proof:

\hskip 0.1cm

\begin{lemma} \label{lemma5} (Channel twirling \cite{bennett1996mixed})

\begin{equation}
\sum_i P_i Q P_i \rho P_i Q' P_i = 0, \text{ if } Q \neq Q' \label{use_prop}
\end{equation}

where $\rho$ is a matrix, $Q$, $Q'$ are two arbitrary general Pauli operators of the same dimension, and $\{P_i\}$ is the set of all general Pauli operators of dimension same as $Q$ and $Q'$.

\end{lemma}

Extra notation has to be introduced first. Vector $\nu$ is used to represent all random secret parameters chosen by Alice throughout the execution of protocol, including $\boldsymbol{r}=\{r_i \}$, $\boldsymbol{\theta}=\{\theta_i \}$, $\boldsymbol{d}=\{d_i\}$ and positions of traps $\boldsymbol{t}=\{t_i \}$. Parameter $p(\nu)$ gives the probability of a particular choice of random secret parameters. Summing over a vector (e.g. $\sum_{\boldsymbol{r}}$) means that we sum over all possible choices for the elements of that vector (e.g. all possible bit-strings of size $N$ for $\boldsymbol{r}$).

For convenience we denote some of the subsystems of the joint Alice-Bob state:
\begin{itemize}
\item System $\mathcal{M}$ contains the qubits of the graph state that participate in the computation and are measured.
\item System $D$ contains all dummy qubits.
\item System $T$ contains all trap qubits.
\item System $O'$ is the fixed output system (therefore all the bottom qubits of the gadget states - which are the only that are not measured).
\item System  $\Delta$ is the system of the measurement angles send by Alice to Bob, where each angle is represented by three qubits in the computational basis.
\item System $B$ is Bob's private system, assumed to be initially in the blank state. 
\end{itemize}


\begin{proof}[\textbf{Proof Sketch of Theorem \ref{theorem1}} (Preparation of verifiable states by the localising protocol.)]

We follow a single index convention to enumerate the qubits participating in the graph where $N$ is the total number of qubits and the last $n'$ qubits are the qubits which are not measured (system $O'$). While in Protocol \ref{prot:vubqc_notraps} system $O'$ remains to the hands of Bob, to be used for the next phase of the hybrid protocol, to prove the property of Theorem \ref{theorem1} we have to assume extra steps of Alice receiving and decoding the output state. As it will be made clear in the next section, the same property is true in the case Bob applies an extra Clifford circuit, which is needed in FK-ABE hybrid protocol, before returning the state to Alice.

Honest computation includes global entangling operator $E_G$ (Step \ref{step5}) on the received state $\ket{M(\boldsymbol{\theta},\boldsymbol{d},\boldsymbol{t})}$ (all qubits sent at Step \ref{step4}) and for each $1 \leq i \leq N-n'$ a measurement, described in Step \ref{step6d}, with outcome $b_i \in \{0,1\}$. This measurement can be analysed into unitary part $H_{i} R_{i}(\delta_i)$, where $H_i$ is the Hadamard gate and $R_{i}(.)$ is rotation around $z$-axis on qubit $i$ which depends on measurement angles $\boldsymbol{\delta}(\boldsymbol{\theta},\boldsymbol{r},\boldsymbol{b})$ (where $\boldsymbol{b} = (b_i)_i$),  and a Pauli $Z$ measurement on qubit $i$ (projectors $\ket{b_i} \bra{b_i}$). Without loss of generality we can represent any dishonest behaviour of Bob at any step as applying the correct unitary operators and then an arbitrary unitary attack operator which applies also on his private system assumed to be in state $\ket{0}^{\otimes B}$.

After trivial commutations, the output of Protocol \ref{prot:vubqc_notraps}, with extra steps of Alice receiving system $O'$ and applying decoding map $\mathcal{D}$ for the FK protocol QECC and final Pauli corrections $C(\boldsymbol{r},\boldsymbol{b})$, which include output system corrections $C_{O'}$  and trap results corrections by $r_t$, can be written as (note that we omit the dependences of each operator in the formula for the sake of brevity):

\begin{eqnarray}
\rho_{out}=Tr_{B,\mathcal{M},\Delta,D} (\sum_{\boldsymbol{b}}  \sum_{\nu} p(\nu)   C  \mathcal{D}  (   \ket{\boldsymbol{b}}\bra{\boldsymbol{b}} U (   H_{N-n'} R_{N-n'}(.) \ldots H_1 R_1(.) E_G ( \nonumber \\ \ket{M}\bra{M} 
 \otimes  \ket{\boldsymbol{\delta}}\bra{\boldsymbol{\delta}} ) E_G R_1(.)^{\dagger}  H_1 \ldots R_{N-n'}(.)^{\dagger} H_{N-n'}   \otimes \ket{0}\bra{0}^{\otimes B} ) 
     U^{\dagger} \ket{\boldsymbol{b}}\bra{\boldsymbol{b}}   )    C^{\dagger} )
\end{eqnarray}

where $U$ is the global (commuted) unitary attack operator that is chosen by Bob and applies on system $\mathcal{M} \times O' \times D \times \Delta \times B$.

Tracing out system $B$, attack $U$ ($U^{\dagger}$) is transformed to a Completely Positive Trace Preserving (CPTP) map acting on the remaining system. We can write the CPTP map in the Kraus decomposition and express the Kraus operators in the Pauli basis (all possible tensor products of single qubit Pauli operators). Moreover, applying the part of the honest unitary operation for each trap (entangling with its neighbour dummy qubits and $H_t R_t(\delta_t)$) results in the system of traps $T$ being disentangled from the rest. The remaining unitary operators are denoted by $\mathcal{P}' \equiv H_{N-n'} R_{N-n'}(.) \ldots  H_1 R_1(.) E_G'$ and the remaining system that they apply to is denoted by $\ket{M'}$.

\begin{eqnarray}
\rho_{out}=Tr_{\mathcal{M},\Delta,D} (\sum_{\boldsymbol{b}}  \sum_{\nu} p(\nu) \sum_{k,u,v,u',v'}  a_{k,u,v} a_{k,u',v'}^*   C_{O'}   \mathcal{D} (  \ket{\boldsymbol{b}'}\bra{\boldsymbol{b}'} P_{u|{i: \forall j, i \neq t_j}}  \otimes P_{v} \nonumber \\ (   \mathcal{P}' (  \ket{M'}\bra{M'} 
 \otimes  \ket{\boldsymbol{\delta}}\bra{\boldsymbol{\delta}} ) \mathcal{P}'^{\dagger}  ) 
      P_{u'|{i: \forall j, i \neq t_j}} \otimes P_{v'} \ket{\boldsymbol{b}'}\bra{\boldsymbol{b}'}   )   C_{O'}^{\dagger} \nonumber \\ \bigotimes_{i}   \ket{b_{t_i}+r_{t_i}}\bra{b_{t_i}} P_{u|{t_i}} \ket{r_{t_i}} \bra{r_{t_i}} P_{u'|{t_i}} \ket{b_{t_i}}\bra{b_{t_i}+r_{t_i}}))
\end{eqnarray}

where
\begin{itemize}
\item $ a_{k,u,v}$ are complex numbers, with $\sum_{k,u,v} |a_{k,u,v}|^2 = 1$.
\item $P_{u}$ (and $P_{u'}$) ranges over all tensor products of Pauli operators on the system that is trapified: all qubits of system $\mathcal{M} \times D \times T$ except the qubits that belong to the gadget and are measured (first row of each gadget).
\item $P_{v}$ (and $P_{v'}$) ranges over all tensor products of Pauli operators on the system that is not trapified: all qubits that belong to gadget systems and the measurement angle system $\Delta$.
\item $P_{u|i}$ is the $i$-th Pauli element of $P_{u}$.
\item $\boldsymbol{b}'$ is the vector that is generated from $\boldsymbol{b}$ by removing elements $\{b_{t_i} \}$.
\end{itemize}

So far, the analysis followed similar techniques to verifiability proof of the original FK protocol. However the following steps have to be introduced to take care of the issues of using the extra gadget state and not having traps on the final output state (therefore proving the relaxed verification requirement).

Summing over random parameters $\boldsymbol{\theta}$, $\boldsymbol{d}$ and $\boldsymbol{r}$ and tracing over systems $\Delta$, $D$ and $\mathcal{M}$ respectively enables us, with the aid of Lemma $\ref{lemma5}$, to eliminate all terms where $P_{u|i} \neq P_{u'|i}$ and $P_{v|i} \neq P_{v'|i}$ which reduces the attack of Bob on the remaining state $\mathcal{P}^{(3)}  \ket{M^{(3)}}$ (where all dependences on these random parameters is eliminated - corrections on the output are also updated accordingly to $C'_{O'}$)  to  a convex combination of Pauli operations.

\begin{eqnarray}
\rho_{out}= \sum_{\boldsymbol{b}}  \sum_{\boldsymbol{t}} p(\boldsymbol{t}) \sum_{k,u,v}  |a_{k,u,v}|^2   C'_{O'}  \mathcal{D}  ( \bra{\boldsymbol{b}'} P_{u|{i: \forall j, i \neq t_j}}  \otimes P_{v|O''} \nonumber \\ (   \mathcal{P}^{(3)} (  \ket{M^{(3)}}\bra{M^{(3)}}  ) \mathcal{P}^{(3)\dagger}  ) 
      P_{u|{i: \forall j, i \neq t_j}} \otimes P_{v|O''} \ket{\boldsymbol{b}'} )   C'_{O'}  \bigotimes_{i}   P_{u|{t_i}} \ket{0} \bra{0} P_{u|{t_i}} \label{equation_sum}
\end{eqnarray}

where $O''$ is the system that contains the qubits of all gadgets.

We split the terms of the sum over the attack operators $P_u$ into the terms which are deterministically corrected by $\mathcal{D}$ and the terms that are not corrected by $\mathcal{D}$ but with high probability will have an effect  on the trap measurement outcomes signifying that an attack has happened.

Specifically, we first examine state $\sigma_1$ which comes from $\rho_{out}$ by restricting summation to $u \in \mathcal{S}_1$, where $\mathcal{S}_1$ is the set of tensor products of Pauli operators that contain $\leq d$ elements with Pauli $X$ component ($d$ is the number of correctable errors of the code). Attacks $P_{v|O''}$ on the output gadget can be rewritten as Pauli attacks on the output state $O'$ that are not correlated on the state itself. This is possible only because Pauli attacks on the last measurement layer of MBQC can be translated to corrections on the output, which is not the case for attacks on any layer. Moreover, attacks $P_{u|{i: \forall j, i \neq t_j}}$ will be perfectly corrected when applying $\mathcal{D}$, since their footprint is $\leq d$, giving correct output state $\ket{\psi_{c}}$. Separating the attacks to those who have an effect on the trap system, causing Alice to set her indicator bit to $\ket{REJ}$, and those who do not touch any trap (and therefore Alice accepts) we get:

\begin{eqnarray}
\sigma_1 =  \sum_{i}  p'_i   P_{i}  \ket{\psi_{c}}\bra{\psi_{c}})   P_{i}  
  \otimes    \ket{ACC} \bra{ACC}  + \sum_{j} p''_j P_{j}  \ket{\psi_{c}}\bra{\psi_{c}}   P_{j}  
  \otimes    \ket{REJ} \bra{REJ} \label{eq8}
\end{eqnarray} 

where $P_{i}$ ($P_{j}$) range over all tensor products of Pauli operators on the output system $O$ and $p'_i,p''_i$ are probabilities, with $\sum_{i} (p'_i + p''_i)  = \sum_{k,u \in \mathcal{S}_1,v} | a_{k,u,v} |^2$.

Let state $\sigma_2$ include remaining terms of the summation over $u$ in Equation \ref{equation_sum}, where $u \in \mathcal{S}_2$ corresponds to $P_u$ where $> d$ Pauli elements have Pauli $X$ component. 

We calculate the probability the state collapses to the `incorrect' subspace $P_{\perp}=I - \ket{\psi_{c}}\bra{\psi_{c}}$ and all the trap measurement outcomes returned are correct. Since Bob has to guess the output of the trap measurement at, at least, $d$ positions, where the traps can be placed with probability that depends on the amount of traps in the graph, the probability of cheating successfully is exponentially decaying in $d$.

\begin{eqnarray}
Tr(P_{\perp} \bigotimes_{i}  \ket{0}_{i} \bra{0}_{i} \sigma_2 ) \leq \sqrt{ (1/c)^d}
\end{eqnarray}

where $c>1$ is a constant that depends on the ratio of total qubits and traps in the graph. Therefore state $\sigma_2$ can be written as:

\begin{eqnarray}
\sigma_2 \approx_{ \epsilon} p_1   \ket{\psi_{c}}\bra{\psi_{c}} \otimes \ket{ACC} \bra{ACC} 
+ p_2  \rho \otimes \ket{REJ} \bra{REJ} \label{eq9}
\end{eqnarray}

where $\epsilon = \sqrt{ (1/c)^d}$, $p_1$ and $p_2$ are probabilities with $p_1+p_2 = \sum_{k,u\in \mathcal{S}_2,v} | a_{k,u,v} |^2$ and $\rho$ is a density matrix.

Summing again the terms $\sigma_1$ and $\sigma_2$, theorem is satisfied for $\epsilon =  \sqrt{ (1/c)^d}$.

\end{proof}

\section{Hybrid Protocol}

The proposed FK-ABE hybrid protocol combines all the techniques discussed so far, including the partitioning of the high complexity resource state of the FK protocol to sub-graphs, each used for the preparation of a separate encoded state for the ABE protocol, thus reducing the quantum and classical communication complexity.

The protocol, described here as Protocol \ref{prot:hybrid_ver}, is based on a $d$-level version of Protocol \ref{prot:vubqc_notraps}, for $d$ odd prime, which corresponds to the following changes:

\begin{itemize}
\item The state is entangled by applying generalized $cZ$ gates.
\item Alice's states $\ket{+_{\theta _{i}}}$ are replaced by states $T^{c_i} S^{b_i} Z^{a_i} \frac{1}{\sqrt{d}}(\ket{0}+ \ldots + \ket{d-1})$, where $T$ are generalized `$\pi/8$' gates, $S$ generalized phase gates, $Z$ generalized Pauli $Z$ gates and $a_i,b_i,c_i \leftarrow_R \mathbf{F}_d$ (except qutrits, where $T$ gate is replaced by $T_3$ gate and $c_i \leftarrow_R \mathbf{F}_9$).
\item Dummy states are generalized computational basis states $\ket{d_i}$, $d_i \leftarrow_R \mathbf{F}_d$
\item Measurements on vector $(a'_i,b'_i,c'_i) \in \mathbf{F}_d^3$ correspond to generalized measurements on basis $\{Z^j T^{c'_i} S^{b'_i} Z^{a'_i} \frac{1}{\sqrt{d}}(\ket{0}+ \ldots + \ket{d-1})\}_{j=0}^{d-1}$. Measurements vectors are corrected to incorporate generalized Pauli $X$ and $Z$ corrections \cite{zhou2003quantum},\cite{hall2005cluster} according to flow dependencies.
\item Alice cancels the pre-rotation by adding $(a_i,b_i,c_i)$ on her (corrected) measurement vector. In the measurement vector she sends to Bob, extra term $(r_i,0,0)^T$, $r_i \leftarrow_R \mathbf{F}_d$ is added to randomize the output of Bob's measurement. She corrects the measurement result she receives from Bob by adding $r_i$.
\end{itemize}

Correctness of the $d$-level version of the FK protocol follows the same argument of the correctness of qubit case, applying $d$-level MBQC \cite{zhou2003quantum},\cite{hall2005cluster} and noticing that the $d$-level dummy qudits have the same effect as in the qubit case. Universality comes from the fact that we are able to perform a universal set of gates for $d$-level computation: $\{cX, F,S,T,Z\}$ \citep{howard2012qudit}, with the special case of qutrits where $T$ becomes $T_3$.

\begin{algorithm}

\caption{ FK-ABE Hybrid Protocol  \label{prot:hybrid_ver}}

\hskip 0.2cm 

\textbf{Alice's input.} Description of a computation in the Gate Teleportation model based on generalized Toffoli states. The input is set to be the Fourier basis state of $n$ qudits: $\ket{+_0}^{\otimes n}$, where $\ket{+_0}=\frac{1}{\sqrt{d}}\sum_a \ket{a}$. The total number of gates is denoted by $t$. Protocol can be extended to admit arbitrary quantum input using the same methods as in the FK protocol.

\hskip 0.2cm 

\textbf{Alice's output.} The result of measurement of the quantum output of the circuit and a bit indicating if the result is accepted or not.

\hskip 0.2cm 

\textbf{The protocol}
\begin{enumerate}
\item Alice chooses a single random sign key (of size equal to the size of a codeword) that will be used to encode all inputs (including Toffoli states) according to the signed-polynomial QECC.
\item A $d$-level Fault Tolerant QECC is selected for amplification of the detection probability of the FK protocol.
\item Alice has to remotely prepare the Toffoli states and to encode both the inputs and the Toffoli states according to the ABE protocol QECC randomized by the selected sign key. For each encoded state she selects a separate graph state that has the properties described in Protocol \ref{prot:vubqc_notraps}. Therefore, the size of each sub-graph depends only on the security parameters of the protocol and not on the size of the computation $t$. 
\item Alice and Bob encode all inputs (including Toffoli states) by executing Protocol \ref{prot:vubqc_notraps} adapted for $d$ level systems. The output of Bob is encoded by the QECC used for amplification of the FK protocol and the final quantum one time pad, which depends on the secret parameters of the FK protocol and will referred to as the Pauli key. Alice holds an indicator bit set on accept or reject depending on the outcome of the traps.
\item Bob applies the Clifford circuit for decoding the QECC used for amplification of the FK protocol and Alice updates her Pauli keys accordingly.
\item Alice and Bob perform the logical operators that correspond to the desired computation on the encoded by polynomial QECC state as described in the ABE protocol. For each application of logical Toffoli gate, Bob sends measurement results to Alice and Alice calculates the actual correction to be performed on the state and sends it to Bob who performs it.
\item Bob measures the output qudits in the computational basis and returns the measurement results to Alice.
\item Alice applies the detection and decoding procedure of the signed polynomial QECC and sets a second indicator bit accordingly.
\item Alice accepts if both indicator bits of the FK and ABE protocol phase of the protocol are set to accept, otherwise she rejects.
\end{enumerate}

\end{algorithm}

A sketch of the proof of verifiability of Protocol \ref{prot:hybrid_ver} is given here, while a detailed proof is presented in Appendix \ref{app_b}.

We use the same notation as in the proof of Theorem \ref{theorem1} to denote the different subsystems, however there is a new subsystem denoted by $\mathcal{L}$ that contains the corrections sent from Alice to Bob for the Toffoli gate implementation at the ABE protocol phase. System $\mathcal{M}$ still contains the qudits of the graph state that participate in the computation and are measured during the application of the FK protocol stage but not the qudits that are measured during the ABE protocol phase. 


\begin{proof}[\textbf{Proof Sketch of Theorem \ref{theorem2}} (Verifiability of FK-ABE hybrid protocol)]

We employ similar techniques to those in Theorem \ref{theorem1} to write the state of Alice, before applying the detection mechanism of the ABE protocol, as a verifiable state, for any choice of the random sign key and without leaking information about it.

The extra elements that FK-ABE hybrid protocol introduces, compared to the localising protocol, include the measurements needed for the implementation of the Toffoli gates, with outcomes $\widetilde{b}_i \in  \mathbf{F}_d$ sent from Bob to Alice and corrections $\widetilde{r}_i \in  \mathbf{F}_d$ sent from Alice to Bob. The Clifford circuit that implements the logical public circuit of the ABE protocol is denoted by $\mathcal{C}$ and applies on the decoded for the FK protocol QECC output system $O$ and on the corrections system $\mathcal{L}$. Final Pauli corrections $C''(\boldsymbol{r},\boldsymbol{b},\boldsymbol{\widetilde{b}})$ for Alice, here, take also into account the update due to the application of $\mathcal{C}$ and  include output system corrections $C''_{O'}$  and trap results corrections by $r_t$. Note that the qubit gates have been replaced by their $d$-level analogues. To highlight the analogy with the qubit case we keep the same notation of the measurement vectors: $\boldsymbol{\theta} , \boldsymbol{\phi} $ and $\boldsymbol{\delta}$, noting that they are dit instead of bit triples. We can, therefore, write  Alice's final output before the detection mechanism of the ABE protocol phase as:

\begin{eqnarray}
\rho_{out}=Tr_{B,\mathcal{M},\Delta,D,\mathcal{L}} (\sum_{\boldsymbol{b},\boldsymbol{\widetilde{b}}}  \sum_{\nu} p(\nu)   C'' (   \ket{\boldsymbol{b}}\bra{\boldsymbol{b}} \otimes \ket{\boldsymbol{\widetilde{b}}}\bra{\boldsymbol{\widetilde{b}}} U \mathcal{C} \mathcal{D} (   F_{N-n'} R_{N-n'}(.) \ldots F_1 R_1(.) E_G ( \nonumber \\ \ket{M}\bra{M} 
 \otimes  \ket{\boldsymbol{\delta}}\bra{\boldsymbol{\delta}} ) E_G R_1(.)^{\dagger}  F_1^{\dagger} \ldots R_{N-n'}(.)^{\dagger} F_{N-n'}^{\dagger}   \otimes \ket{0}\bra{0}^{\otimes B} \otimes \ket{\widetilde{\boldsymbol{r}}}\bra{\widetilde{\boldsymbol{r}}})
  \mathcal{D}^{\dagger}  \mathcal{C}^{\dagger} U^{\dagger} \ket{\boldsymbol{b}}\bra{\boldsymbol{b}} \otimes \ket{\boldsymbol{\widetilde{b}}}\bra{\boldsymbol{\widetilde{b}}}  ) C'' )
\end{eqnarray}

where $U$ represents global (commuted) Bob's attack that in this case applies also on system $\mathcal{L}$.

Similarly to proof of Theorem \ref{theorem1} Alice's output can be rewritten as:

\begin{eqnarray}
\rho_{out}=Tr_{\mathcal{M},\Delta,D,\mathcal{L}} (\sum_{\boldsymbol{b},\boldsymbol{\widetilde{b}}}  \sum_{\nu} p(\nu) \sum_{k,u,v,u',v'}  a_{k,u,v} a_{k,u',v'}^*   C''_{O'} (   \ket{\boldsymbol{b}'}\bra{\boldsymbol{b}'}  \otimes \ket{\boldsymbol{\widetilde{b}}}\bra{\boldsymbol{\widetilde{b}}} P_{u|{i: \forall j, i \neq t_j}}  \otimes P_{v}  \nonumber \\ \mathcal{C}  \mathcal{D}(   \mathcal{P}' (  \ket{M'}\bra{M'} 
 \otimes  \ket{\boldsymbol{\delta}}\bra{\boldsymbol{\delta}} ) \mathcal{P}'^{\dagger}  ) \mathcal{D}^{\dagger} \mathcal{C}^{\dagger} 
      P_{u'|{i: \forall j, i \neq t_j}} \otimes P_{v'} \ket{\boldsymbol{b}'}\bra{\boldsymbol{b}'}  \otimes \ket{\boldsymbol{\widetilde{b}}}\bra{\boldsymbol{\widetilde{b}}}  ) C''^{\dagger}_{O'} \nonumber \\ \bigotimes_{i}   \ket{b_{t_i}+r_{t_i}}\bra{b_{t_i}} P_{u|{t_i}} \ket{r_{t_i}} \bra{r_{t_i}} P_{u'|{t_i}} \ket{b_{t_i}}\bra{b_{t_i}+r_{t_i}}))
\end{eqnarray}

where the only difference from the proof of Theorem \ref{theorem1} is that the attacks $P_{u}$ (and $P_{u'}$), $P_{v}$ (and $P_{v'}$) are tensor products of  $d$-level Pauli operators  and that $P_{v}$ (and $P_{v'}$) applies also on system $\mathcal{L}$.

Again, summing over random parameters $\boldsymbol{\theta}$, $\boldsymbol{d}$ and $\boldsymbol{r}$ and tracing over systems $\Delta$, $D$ and $\mathcal{M}$ respectively enables us to reduce the attack on these systems to a mixture of Pauli operations. Special care has to be taken for the attack on system $\mathcal{L}$ which can be reduced to Pauli operators by exploiting the structure of the Toffoli implementation circuit, where all logical measurements have uniformly random results, and, therefore, an extra logical Pauli X randomization can be extracted. Tracing out the system $\mathcal{L}$ as well results to:

\begin{eqnarray}
\rho_{out}= \sum_{\boldsymbol{b},\boldsymbol{\widetilde{b}}}  \sum_{\boldsymbol{t}} p(\boldsymbol{t}) \sum_{k,u,v}  |a_{k,u,v}|^2   C^{(3)}_{O'}  ( \bra{\boldsymbol{b}'} \otimes \ket{\boldsymbol{\widetilde{b}}}\bra{\boldsymbol{\widetilde{b}}} P_{u|{i: \forall j, i \neq t_j}}  \otimes P_{v|O''}  \nonumber \\ (  \mathcal{C} \mathcal{D}( \mathcal{P}^{(3)} (  \ket{M^{(3)}}\bra{M^{(3)}}  ) \mathcal{P}^{(3)\dagger}) \mathcal{D}^{\dagger} \mathcal{C}^{\dagger}  ) 
      P_{u|{i: \forall j, i \neq t_j}} \otimes P_{v|O''} \ket{\boldsymbol{b}'}  \otimes \ket{\boldsymbol{\widetilde{b}}}\bra{\boldsymbol{\widetilde{b}}}  ) C^{(3)\dagger}_{O'}  \bigotimes_{i}   P_{u|{t_i}} \ket{0} \bra{0} P_{u|{t_i}}  \label{eq12}
\end{eqnarray}

The next step is the splitting of the internal sum (over $u$) into two parts, yielding two sub-normalized states $\sigma_1$ and $\sigma_2$. In $\sigma_1$ we restrict the index $u$ to the set $S_1$, which contains all attacks which are corrected by the FK protocol QECC, with parameter $d_1$, and in $\sigma_2$ to $u \in \mathcal{S}_2$ which contains attacks with Pauli X footprint $> d_1$ and therefore will activate at least $d_1$ traps when averaging over all positions for the traps. The only difference here from the proof of Theorem \ref{theorem1} is the existence of the public circuit $\mathcal{C}$, but since it applies only on system $O$ it does not affect the probabilities of detecting the traps and since it is Clifford the Pauli corrections commute with it and get updated secretly by Alice. Again:

\begin{eqnarray}
\sigma_1 =  \sum_{i}  p'_i   P_{i} \ket{\psi_{c}'}\bra{\psi_{c}'}   P_{i}  
  \otimes    \ket{ACC_1} \bra{ACC_1}  + \sum_{j} p''_j P_{j} \ket{\psi_{c}'}\bra{\psi_{c}'}    P_{j}  
  \otimes    \ket{REJ_1} \bra{REJ_1} \label{eq_s1}
\end{eqnarray}

where $\ket{\psi_{c}'}\bra{\psi_{c}'}$ is the correct state after the ABE protocol phase is performed and the quantum one time pad with the Pauli key is undone. Also $P_{i}$ ($P_{j}$) range over all tensor products of $d$-level Pauli operators on the output system $O$ and $p'_i,p''_i$ are probabilities, with $\sum_{i} (p'_i + p''_i)  = \sum_{k,u \in \mathcal{S}_1,v} | a_{k,u,v} |^2$.

Also, since the same counting arguments hold for the $d$-level traps on the measured system:

\begin{eqnarray}
\sigma_2 \approx_{ \epsilon_1} p_1   \ket{\psi_{c}'}\bra{\psi_{c}'}  \otimes \ket{ACC_1} \bra{ACC_1} 
+ p_2  \rho \otimes \ket{REJ_1} \bra{REJ_1} \label{eq_s2}
\end{eqnarray}

where $\epsilon_1 = \sqrt{ (1/c)^{d_1}}$, $p_1+p_2 = \sum_{k,u\in \mathcal{S}_2,v} | a_{k,u,v} |^2$ and $\rho$ is a density matrix.

By adding the states described in Equations \ref{eq_s1} and \ref{eq_s2} and by contractiveness of the ABE protocol detection procedure that Alice applies to check if the state is in the valid space of the signed polynomial QECC, theorem is satisfied for $\epsilon_1 =  \sqrt{ (1/c)^{d_1}}$ and  $\epsilon_2 =  (1/2)^{d_2}$, where $d_2$ is the degree of the polynomials in the QECC used in the ABE protocol.

\end{proof}

In the FK-ABE hybrid protocol the quantum requirement for Alice is to prepare single qudit states of the form  $T^{c_i} S^{b_i} Z^{a_i} \ket{+_0}$, for  $a_i,b_i,c_i \leftarrow_R \mathbf{F}_d$, where $d$ is an odd prime. The dimension $d$ of each system is $O(1/log(\epsilon))$. Alice has to send to Bob, at the beginning of the protocol before getting the computation description, $O(n \text{Polylog}(\frac{1}{\epsilon}))$ separable single qudit states,  where $n$ is the size of the computation. After, Alice and Bob have to exchange only classical information in $O(n \text{Polylog}(\frac{1}{\epsilon}))$ rounds, where each round includes a constant size message sent from Bob to Alice and a constant size message from Alice to Bob.
 
\section*{Acknowledgements}

The authors would like to thank Petros Walden for the helpful discussions. TK also wants to thank Telecom ParisTech for its hospitality during his visit there, where a portion of this work had been accomplished. VD was funded by the EPSRC Doctoral Prize Fellowship during the development of this work.

\bibliographystyle{unsrt}

\appendix

\section{Proof of Theorem \ref{theorem1}} \label{app_a}

Single index notation is followed to enumerate the qubits participating in the graph where $N$ is the total number of qubits and the last $n'$ qubits are the encoded output qubits (system $O'$). Each measurement performed at Step \ref{step6d} of the protocol is analysed into a unitary part and a Pauli $Z$ measurement. Without loss of generality we can represent any dishonest behaviour of Bob at any step as applying the correct unitary operators and then an arbitrary unitary attack operator.

The output that Alice produces after she receives system $O'$ from Bob and applies the decoding operators, averaged over all random parameters is:

\begin{eqnarray}
\rho_{out}=Tr_{B,\mathcal{M},\Delta,D} (\sum_{\boldsymbol{b}}  \sum_{\nu} p(\nu)   C \mathcal{D}(  
\ket{b_{N-n'}}\bra{b_{N-n'}} U_{N-n'} H_{N-n'} R_{N-n'}(\delta_{N-n'}) (\ket{\delta_{N-n'}} \bra{\delta_{N-n'}} \otimes \ldots \nonumber \\
 \ket{b_{k}}\bra{b_{k}} U_{k} H_k R_k(\delta_k) (\ket{\delta_k} \bra{\delta_k} \otimes \nonumber \\
\ldots \ket{b_{1}}\bra{b_{1}} U_{1} H_1 R_1(\delta_1) (\ket{\delta_1}\bra{\delta_1} \otimes E_G \ket{M} \bra{M} \otimes \ket{0}\bra{0}^{\otimes B} E_G) R_1(\delta_1)^{\dagger} H_1 U_{1}^{\dagger} \ket{b_{1}} \bra{b_{1}} \ldots \nonumber \\
) R_k(\delta_k)^{\dagger} H_k U_{k}^{\dagger}  \ket{b_{k}}\bra{b_{k}} \ldots ) R_{N-n'}(\delta_{N-n'})^{\dagger} H_{N-n'} U_{N-n'}^{\dagger} \ket{b_{N-n'}}\bra{b_{N-n'}} )  C^{\dagger})
\end{eqnarray}

where:

\begin{itemize}
\item $\ket{M(\boldsymbol{\theta},\boldsymbol{d},\boldsymbol{t})}$  is the state that Bob receives after Step \ref{step4} of the protocol (input, auxiliary qubits, dummy qubits).
\item $E_G$ is the global entangling operator that Bob applies at step 5 of the protocol.
\item $\delta(\boldsymbol{\theta},\boldsymbol{r},\boldsymbol{b})$ is the vector of $\delta_i(\boldsymbol{\theta},\boldsymbol{r},\boldsymbol{b})$ which are measurement angles that Alice sends to Bob at iteration $i$ of step 5 of the protocol .
\item $R_i(\delta_i)$ is rotation of qubit $i$ around $z$-axis by angle $\delta_i$.
\item $H_i$ is a Hadamard gate applied on qubit $i$.
\item $U_{i}$ is the unitary attack of Bob with index $i$ and applies on qubits $\geq i$, private Bob's system and all measurement angles.
\item $\mathcal{D}$ is the map that Alice applies to decoding for the QECC used in the Fault Tolerant procedure. 
\item $C(\boldsymbol{r},\boldsymbol{b})$ includes the Pauli correction $C_{O'}$ that Alice performs to the output system $O'$ and the corrections to trap measurement results.
\end{itemize}

By commuting all measurement angles (trivially) and all attack unitary operators (by observing that they trivially commute with all measurements $\leq i$) and merge them into a new unitary $U$ that applies on the whole of the system we get (see also Figure \ref{figure1}):

\begin{figure}[h!]
  \centering
      \includegraphics[width=1\textwidth]{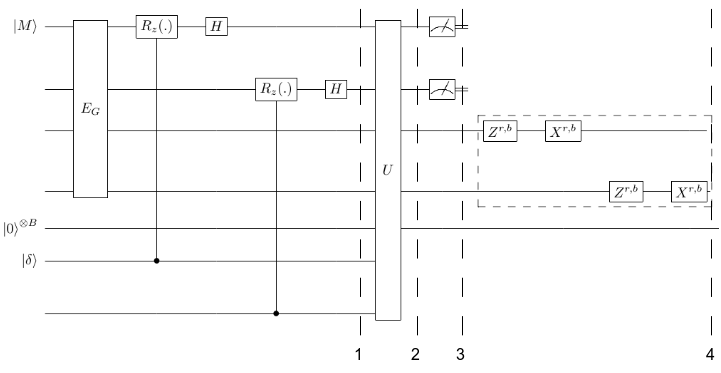}
  \caption{Analysis of attacks in the localising protocol. Part inside dashed box is applied by Alice.} \label{figure1}
\end{figure} 

\begin{eqnarray}
\rho_{out}=Tr_{B,\mathcal{M},\Delta,D} (\sum_{\boldsymbol{b}}  \sum_{\nu} p(\nu)   C  \mathcal{D}  (   \ket{\boldsymbol{b}}\bra{\boldsymbol{b}} U (   H_{N-n'} R_{N-n'}(.) \ldots H_1 R_1(.) E_G ( \nonumber \\ \ket{M}\bra{M} 
 \otimes  \ket{\boldsymbol{\delta}}\bra{\boldsymbol{\delta}} ) E_G R_1(.)^{\dagger}  H_1 \ldots R_{N-n'}(.)^{\dagger} H_{N-n'}   \otimes \ket{0}\bra{0}^{\otimes B} ) 
     U^{\dagger} \ket{\boldsymbol{b}}\bra{\boldsymbol{b}}   )    C^{\dagger} ) \label{a1}
\end{eqnarray}

where $U$ is a unitary attack operator that is chosen by Bob and applies on all graph qubits, private Bob's system and all measurement angles.

The trap system is not entangled with the rest when the honest computation is applied. We can separate the terms that apply on the trap (remember that the trap is placed among the measured qubits). Applying the entangling operators between the traps and their neighbours, which are always dummy qubits does not have any effect other than undoing the pre-rotation of the trap (for the case the dummy is selected to be $\ket{1}$). Applying $H_t R_t(\delta_t)$ on each trap, where $\delta_t=\theta_t + r_t \pi$, results in getting state $\ket{r_t}$.

Therefore, it holds that:

\begin{eqnarray}
 \mathcal{P} (  \ket{M}\bra{M} 
 \otimes  \ket{\boldsymbol{\delta}}\bra{\boldsymbol{\delta}} ) \mathcal{P}^{\dagger} =   \mathcal{P}' (  \ket{M'}\bra{M'}
 \otimes  \ket{\boldsymbol{\delta}}\bra{\boldsymbol{\delta}} ) \mathcal{P}'^{\dagger}  \bigotimes_{\boldsymbol{i}}     \ket{r_{t_i}} \bra{r_{t_i}} \label{a2}
\end{eqnarray}

where
\begin{itemize}
\item $\mathcal{P}=H_{N-n'} R_{N-n'}(.) \ldots  H_{t_i} R_{t_i}(.) \ldots  H_1 R_1(.) E_G$ is the correct unitary operation in Figure $\ref{figure1}$.
\item $\mathcal{P}'=H_{N-n'} R_{N-n'}(.) \ldots  H_1 R_1(.) E_G'$, which is produced from $\mathcal{P}$ by omitting all operations applied on traps ($E_G'$ is produced from $E_G$ by omitting all entangling operators between traps and their neighbours) 
\item $\ket{M'}$ is produced from $\ket{M}$ by omitting the trap qubits. 
\end{itemize}

In the next step we express the output state with Bob's private system traced out and his attack decomposed to the Pauli basis. We use the following general property:

For any unitary $U$ and $n$-qubit state $\rho$ it holds that:
\begin{eqnarray}
Tr_B (U (\rho \otimes \ket{0} \bra{0} ^{B} ) U^{\dagger}) =  \sum_{k,u,u'}  a_{k,u} a_{k,u'}^* P_{u} \rho P_{u'}
\end{eqnarray}
where 
\begin{itemize}
\item $ a_{k,u}$ are complex numbers, with $\sum_{k,u} |a_{k,u}|^2 = 1$.
\item $P_{u}$ (and $P_{u'}$) ranges over all general Pauli operators on $n$ qubits.
\end{itemize}

Applying this property to the state in Equation \ref{a1}, by tracing out system $B$, and at the same time applying Equation \ref{a2} to separate the traps we get the following form:

\begin{eqnarray}
\rho_{out}=Tr_{\mathcal{M},\Delta,D} (\sum_{\boldsymbol{b}}  \sum_{\nu} p(\nu) \sum_{k,u,v,u',v'}  a_{k,u,v} a_{k,u',v'}^*   C_{O'}   \mathcal{D} (  \ket{\boldsymbol{b}'}\bra{\boldsymbol{b}'} P_{u|{i: \forall j, i \neq t_j}}  \otimes P_{v} \nonumber \\ (   \mathcal{P}' (  \ket{M'}\bra{M'} 
 \otimes  \ket{\boldsymbol{\delta}}\bra{\boldsymbol{\delta}} ) \mathcal{P}'^{\dagger}  ) 
      P_{u'|{i: \forall j, i \neq t_j}} \otimes P_{v'} \ket{\boldsymbol{b}'}\bra{\boldsymbol{b}'}   )  C^{\dagger}_{O'} \nonumber \\ \bigotimes_{i}   \ket{b_{t_i}+r_{t_i}}\bra{b_{t_i}} P_{u|{t_i}} \ket{r_{t_i}} \bra{r_{t_i}} P_{u'|{t_i}} \ket{b_{t_i}}\bra{b_{t_i}+r_{t_i}}))
\end{eqnarray}

where
\begin{itemize}
\item $ a_{k,u,v}$ are complex numbers, with $\sum_{k,u,v} |a_{k,u,v}|^2 = 1$.
\item $P_{u}$ (and $P_{u'}$) ranges over all general Pauli operators and applies on the system that is trapified: all systems that are measured except those on the output gadget.
\item $P_{v}$ (and $P_{v'}$) ranges over all general Pauli operators and applies on the system that is not trapified: the whole of the gadget system $O''$ and the measurement angle system $\Delta$.
\item $P_{u|i}$ is the $i$-th Pauli element of $P_{u}$.
\item $\boldsymbol{b}'$ is the vector that is generated from $\boldsymbol{b}$ by removing elements $\{b_{t_i} \}$. 
\end{itemize}

In the next part of the proof the attack of Bob will be reduced to a mixture of Pauli operators, utilizing all random parameters except the random position of the trap to cancel the cross Pauli terms that are not the same.

By observing that $R_i(.) (\ket{\psi} \otimes \ket{\delta_i})$ is equivalent mathematically to $R_i(\delta_i) \ket{\psi} \otimes \ket{\delta_i}$, since $\delta_i$ is a classical state we can rewrite 

\begin{eqnarray}
\mathcal{P}'=H_{N-n'} R_{N-n'}(\delta_{N-n'}) \ldots  H_1 R_1(\delta_1) E_G' = \nonumber \\
H_{N-n'} R_{N-n'}(\phi'_{N-n'} + \theta_{N-n'} + r_{N-n'} \pi) \ldots  H_1 R_1(\phi'_{1} + \theta_{1} + r_{1} \pi) E_G' = \nonumber \\
H_{N-n'} R_{N-n'}(\phi'_{N-n'} + r_{N-n'} \pi) \ldots  H_1 R_1(\phi'_{1} + r_{1} \pi) E_G' R_{N-n'}(\theta_{N-n'}) \ldots R_{N-n'}(\theta_{i}) = \nonumber \\
\mathcal{P}''   R_{N-n'}(\theta_{N-n'}) \ldots R_{N-n'}(\theta_{i})
\end{eqnarray}

Operations $R_{N-n'}(\theta_{N-n'}) \ldots R_{N-n'}(\theta_{i})$ when applied to states $\ket{M}$ (or not applying at all on dummies since $\theta$ is 0) they cancel the pre-rotation and the state is updated to a state which does not depend on $\boldsymbol{\theta}$, denoted by $\ket{M'}$.

Now that we have eliminated all dependences on $\boldsymbol{\theta}$ except for states $\ket{\boldsymbol{\delta}}$ we can sum over $\boldsymbol{\theta}$ to get for system $\Delta$ the state $  P_{v|\Delta} \frac{1}{2^{|\Delta|}} I  P_{v'|\Delta}$. For $P_{v|\Delta}=P_{v'|\Delta}$ this has trace 1 and for  $P_{v|\Delta} \neq P_{v'|\Delta}$ it has trace 0, since all Pauli operators except identity are traceless. Therefore we can ignore the attack cross terms applied on the measurement angles system.

Next we consider the dummy qubits. Part of $E_G'$ is applied between them and the computational qubits cancelling the pre-rotation of the latter when $d_i=1$. The remaining entangling operators will be denoted by $E_G''$. Also for the rest of the operation applied on each dummy $i \in D$:

\begin{eqnarray}
\sum_{d_i,r_i} H_i R_i(r_{i} \pi) \ket{d_i} \bra{d_i} R_i(r_{i} \pi) H_i ) = \frac{1}{2} I 
\end{eqnarray}

Again $Tr(  P_{u|D} \frac{1}{2^{|D|}} I  P_{u'|D})$ is 1 for $P_{u|D}=P_{u'|D}$ and 0 for $P_{u|D} \neq P_{u'|D}$ so we can ignore the cross terms of the attack applying on the dummy qubits. Let $\ket{M''}$ denote state $\ket{M'}$ without the traps and the dummy qubits.

The next step is an iteration over all qubits $i$ of the graph state and using summation over each $\boldsymbol{r_i}$ to twirl each attack operator on qubit $f(i)$ respectively, where $f$ is the flow function. We begin from qubit $N$ and follow the  reverse measuring order. We distinguish between tree types of qubits depending on their placement and we apply a different technique for each one:

\begin{itemize}
\item Qubit $i \in O'$: We extract stabilizer $Z_{f^{-1} (i)}^{r_{f^{-1} (i)}} X_{i}^{r_{f^{-1} (i)}}$ from graph $E_G'' \ket{M''}$ (assuming that the final layer qubits are not entangled between them as it is the case in the graphs we consider in this work). This cancels dependence on $r_{f^{-1} (i)}$ on the operation applied on qubit $f^{-1} (i)$ and allows as to sum over $r_{f^{-1} (i)}$ (and $r_i$) to get property (using Lemma \ref{lemma5}):

\begin{eqnarray}
\sum_{r_{f^{-1} (i)},r_i} Z_i^{b_{f^{-1} (j \sim i)}+r_{f^{-1} (j \sim i)}}  X_{i}^{b_{f^{-1} (i)}+r_{f^{-1} (i)}} Z_i^{r_i} P_{v|i} X_{i}^{r_{f^{-1} (i)}} Z_i^{r_i} E_G'' (  \ket{M^{(3)}}\bra{M^{(3)}}   ) \nonumber \\
E_G'' Z_i^{r_i} X_{i}^{r_{f^{-1} (i)}} P_{v'|i}  Z_i^{r_i} X_{i}^{b_{f^{-1} (i)}+r_{f^{-1} (i)}} Z_i^{b_{f^{-1} (j \sim i)}+r_{f^{-1} (j \sim i)}} = 0 \text{ if } P_{v|i} \neq P_{v'|i}
\end{eqnarray}

Note that for the terms that remain ($P_{v|i} = P_{v'|i}$) summing over $r_{f^{-1} (i)},r_i$ eliminates all dependences from $r_{f^{-1} (i)},r_i$ in the formula. Also note that $Z_i^{r_{f^{-1} (j \sim i)}} $  terms and all the $b$ corrections remain.

\item Qubit $i \notin O' \cup I$: We extract stabilizer $Z_{f^{-1} (i)}^{r_{f^{-1} (i)}} X_{i}^{r_{f^{-1} (i)}} Z_{j \sim i, j \neq f^{-1} (i)}^{r_{f^{-1} (i)}}$ from graph $E_G'' \ket{M^{(3)}}$. This cancels dependence on $r_{f^{-1} (i)}$ on the operation applied on qubit $f^{-1} (i)$ and on operations applied on qubits $\{ j \sim i, j \neq f^{-1} (i)\}$ (notice that attacks on qubits $\{ j \sim i, j \neq f^{-1} (i)\}$ have already been twirled because we follow the opposite of flow). By summing the elements that still depend on $r_{f^{-1} (i)}$:

\begin{eqnarray}
\sum_{r_{f^{-1} (i)}} \bra{b_i} P_{u|i} H_i R_i(\phi_i' + r_{i} \pi) \nonumber \\ X_{i}^{r_{f^{-1} (i)}}  E_G'' (  \ket{M^{(3)}}\bra{M^{(3)}}   ) E_G''  X_{i}^{r_{f^{-1} (i)}} R_i(\phi_i' + r_{i} \pi)^{\dagger} H_i   P_{u'|i} \ket{b_i}
\end{eqnarray}

where $\phi_i' = (-1)^{b_{f^{-1}(i)}+r_{f^{-1}(i)}}\phi_{i} + (b_{f^{-1} (j \sim i, j \neq f(i))}+r_{f^{-1} (j \sim i, j \neq f(i))}) \pi$.

This state is equal to

\begin{eqnarray}
\sum_{r_{f^{-1} (i)}} \bra{b_i} Z_{i}^{r_{f^{-1}}} P_{u|i} Z_{i}^{r_{f^{-1}}} H_i R_i(\phi_i'' + r_{i} \pi) \nonumber \\   E_G'' (  \ket{M^{(3)}}\bra{M^{(3)}}   ) E_G''  R_i(\phi_i'' + r_{i} \pi)^{\dagger} H_i Z_{i}^{r_{f^{-1}}}  P_{u'|i}  Z_{i}^{r_{f^{-1}}} \ket{b_i}  
\end{eqnarray}

where $\phi_i'' = (-1)^{b_{f^{-1}(i)}}\phi_{i} + (b_{f^{-1} (j \sim i, j \neq f(i))}+r_{f^{-1} (j \sim i, j \neq f(i))}) \pi$.

To get the above state we used also the fact that we can extract a random $Z_{i}^{r_{f^{-1}}}$ from both $\bra{b_i}$ and $\ket{b_i}$, without affecting the state.

To apply Lemma \ref{lemma5} we also need the random $X$ elements. These are acquired by extracting stabilizer $Z_{i}^{r'_i} X_{f(i)}^{r'_i} Z_{j \sim f(i), j \neq i}^{r'_i}$ from graph $E_G'' \ket{M^{(3)}}$ and also changing variable $b^{\diamond}_i \leftarrow b_i + r'_i$ for $r'_i \leftarrow_R \{0,1\}$. The new terms cancel everywhere except at qubit $i$ so that we get:

\begin{eqnarray}
\sum_{r_{f^{-1} (i)},r'_i} \bra{b^{\diamond}_i} X^{r'_i} Z_{i}^{r_{f^{-1}}} P_{u|i} Z_{i}^{r_{f^{-1}}} X^{r'_i} H_i R_i(\phi_i'' + r_{i} \pi)    E_G'' (  \ket{M^{(3)}}\bra{M^{(3)}}   ) E_G''  \nonumber \\ R_i(\phi_i'' + r_{i} \pi)^{\dagger} 
 H_i X^{r'_i} Z_{i}^{r_{f^{-1}}}   P_{u'|i}  Z_{i}^{r_{f^{-1}}} X^{r'_i} \ket{b^{\diamond}_i} = 0 \text{ if } P_{u|i} \neq P_{u'|i}
\end{eqnarray}

\item Qubit $i \in I$: We use property $ X R(\phi) \ket{+} = e^{i \phi} R(-\phi) \ket{+}$ We extract stabilizer $ X_{1}^{z_0} Z_{f(i)}^{z_0}$ from graph $E_G'' \ket{M^{(3)}}$ (graphs we use have the property that inputs $i$ are only entangled with $f(i)$), where $z_0 \leftarrow \{0,1\}$. Also  extract a random $Z^{z_0}$ from $\bra{b_i}$ and from $\ket{b_i}$ and employing the same trick for generating the $X$ elements that we did in the previous step and then applying Lemma \ref{lemma5} gives us:

\begin{eqnarray}
\sum_{x_0,z_0} \bra{b_i} X^{x_0} Z^{z_0}  P_{u|i} Z^{z_0} X^{x_0} H_i R_i(\phi_i + r_{i} \pi) \nonumber \\   E_G'' (  \ket{M^{(3)}}\bra{M^{(3)}}   ) E_G''   R_i(\phi_i + r_{i} \pi)^{\dagger} H_i   X^{x_0} Z^{z_0}   P_{u'|i}  Z^{z_0} X^{x_0} \ket{b_i} = 0  \text{ if } P_{u|i}  \neq P_{u'|i}
\end{eqnarray}

\end{itemize}

Finally we consider the trap qubits $t_i$:

\begin{eqnarray}
\sum_{r_{t_i},b_{t_i}}   \ket{b_{t_i}+r_{t_i}}\bra{b_{t_i}} P_{u|{t_i}} \ket{r_{t_i}} \bra{r_{t_i}} P_{u'|{t_i}} \ket{b_{t_i}}\bra{b_{t_i}+r_{t_i}}
\end{eqnarray}

By changing variable $b^{\diamond}_{t_i} \leftarrow b_{t_i} + r_{t_i}$

\begin{eqnarray}
\sum_{r_{t_i},b^{\diamond}_{t_i}}   \ket{b^{\diamond}_{t_i}}\bra{b^{\diamond}_{t_i}} X^{r_{t_i}} P_{u|{t_i}} X^{r_{t_i}} \ket{0} \bra{0} X^{r_{t_i}} P_{u'|{t_i}}   X^{r_{t_i}}\ket{b^{\diamond}_{t_i}}\bra{b^{\diamond}_{t_i}} \nonumber \\
= \sum_{r_{t_i},r'_{t_1},b^{\diamond}_{t_i}}   \ket{b^{\diamond}_{t_i}}\bra{b^{\diamond}_{t_i}} Z^{r'_{t_i}} X^{r_{t_i}} P_{u|{t_i}} X^{r_{t_i}} Z^{r'_{t_i}} \ket{0} \bra{0} Z^{r'_{t_i}} X^{r_{t_i}} P_{u'|{t_i}}  
 X^{r_{t_i}} Z^{r'_{t_i}} \ket{b^{\diamond}_{t_i}}\bra{b^{\diamond}_{t_i}} \nonumber \\ = 0 \text{ if } P_{u|t_i}  \neq P_{u'|t_i}
\end{eqnarray}

Thus we have zeroed all terms where $u \neq u$ and $v \neq v$ which gives:

\begin{eqnarray}
\rho_{out}=\sum_{\boldsymbol{b}}  \sum_{\boldsymbol{t}} p(\boldsymbol{t}) \sum_{k,u,v}  |a_{k,u,v}|^2   C'_{O'}  \mathcal{D}  ( \bra{\boldsymbol{b}'} P_{u|{i: \forall j, i \neq t_j}}  \otimes P_{v|O''} \nonumber \\ (   \mathcal{P}^{(3)} (  \ket{M^{(3)}}\bra{M^{(3)}}  ) \mathcal{P}^{(3)\dagger}  ) 
      P_{u|{i: \forall j, i \neq t_j}} \otimes P_{v|O''} \ket{\boldsymbol{b}'} )   C'^{\dagger}_{O'}  \bigotimes_{i}   P_{u|{t_i}} \ket{0} \bra{0} P_{u|{t_i}}  \label{a3}
\end{eqnarray}

where

\begin{itemize}
\item $\ket{M^{(3)}}$ is the tensor product of $\ket{+}$ for all qubits of system $\mathcal{M} \times O'$.
\item $\mathcal{P}^{(3)}$ is the unitary that contains $E_G'$  for $\mathcal{M} \times O'$ and $H_{i} R_{i}((-1)^{b_{f^{-1}(i)}}\phi_{i} + b_{f^{-1} (j \sim i, j \neq f(i))} \pi) $ operators for all qubits of system $\mathcal{M}$.
\item $C'$ are the remaining Pauli corrections on the output system $O'$ (after eliminating the $r$'s) and depend only on $\boldsymbol{b}$. 
\end{itemize}

We split the sum over the attack operators $P_u$ (applying on the trapified system) in Equation \ref{a3} into the operators which are perfectly corrected by the QECC used in the FK protocol and the operators which are not corrected but have a significant footprint to be caught by the trapification procedure with high probability.

Specifically, restricting the summation in the state of Equation \ref{a3} to $u \in \mathcal{S}_1$  where $\mathcal{S}_1$ is the set of general Pauli operators that contain $\leq d$ elements which have a Pauli $X$ component (where $d$ is the number of correctable errors of the code), we have state:

\begin{eqnarray}
\sigma_1 = \sum_{\boldsymbol{b}}  \sum_{\boldsymbol{t}} p(\boldsymbol{t}) \sum_{k,u\in \mathcal{S}_1,v}  |a_{k,u,v}|^2   C'_{O'} \mathcal{D}  \bra{\boldsymbol{b}'} P_{u|{i: \forall j, i \neq t_j}}  \otimes P_{v|O''} \nonumber \\ (   \mathcal{P}^{(3)} (  \ket{M^{(3)}}\bra{M^{(3)}}  ) \mathcal{P}^{(3)\dagger}  ) 
      P_{u|{i: \forall j, i \neq t_j}} \otimes P_{v|O''} \ket{\boldsymbol{b}'} )   C'^{\dagger}_{O'}  \bigotimes_{i}   P_{u|{t_i}} \ket{0} \bra{0} P_{u|{t_i}} 
\end{eqnarray}

Let us separate the terms of the Pauli attack $P_{v|O''}$ on the gadget system  to those applying on the measured qubits, say $P_{v_1}$, and those who apply on the output system $O'$ qubits, say $P_{v_2}$.

The X-component of attack $P_{v_1}$, denoted by $P'_{v_1}$, can be replaced by a $P'_{v_1}$ attack on output system $O'$ (since it affects the X corrections of the output) and the Z-component of attack $P_{v_1}$ does not have any effect on the state. Note that if we wanted to reduce attacks of previous measurement layers to attacks on the output these attacks would have a correlation with the output, but in the case of the last measurement layer this does not happen.

Commuting the output Pauli attacks $P'_{v_1} P_{v_2}$ with the decoding operator $\mathcal{D}$ updates them to a different Pauli operator $P''_{v_1} P'_{v_2}$ :

\begin{eqnarray}
= \sum_{\boldsymbol{b}}  \sum_{\boldsymbol{t}} p(\boldsymbol{t}) \sum_{k,u\in \mathcal{S}_1,v}  |a_{k,u,v}|^2   C'_{O'}   \bra{\boldsymbol{b}'} P_{u|{i: \forall j, i \neq t_j}}  \otimes P''_{v_1} P'_{v_2} \mathcal{D} \nonumber \\ (   \mathcal{P}^{(3)} (  \ket{M^{(3)}}\bra{M^{(3)}}  ) \mathcal{P}^{(3)\dagger}  ) 
      P_{u|{i: \forall j, i \neq t_j}} \otimes \mathcal{D}^{\dagger} P'_{v_2} P''_{v_1}  \ket{\boldsymbol{b}'} )   C'^{\dagger}_{O'}  \bigotimes_{i}   P_{u|{t_i}} \ket{0} \bra{0} P_{u|{t_i}} 
\end{eqnarray}

Since $P_u \in \mathcal{S}_1$ attack $P_{u|{i: \forall j, i \neq t_j}}$ and the QECC perfectly corrects $d$ errors, the attack will not  have any effect on the correct (quantum one-time-padded) outcome after applying the measurements and decoding.

\begin{eqnarray}
=\sum_{\boldsymbol{b}}  \sum_{\boldsymbol{t}} p(\boldsymbol{t}) \sum_{k,u \in \mathcal{S}_1,v}  |a_{k,u,v}|^2      C'_{O'} P''_{v_1} P'_{v_2}  (   C'_O \ket{\psi_{c}}\bra{\psi_{c}} C'_{O'} ) P'_{v_2} P''_{v_1}  C'^{\dagger}_O   )   \bigotimes_{i}   P_{u|{t_i}} \ket{0} \bra{0} P_{u|{t_i}} 
\end{eqnarray}

We can further eliminate $C'_{O'}$ by commuting through output attacks.

By separating the terms that leave all traps untouched and the rest and tracing out $O'\setminus O$ we get: 

\begin{eqnarray}
\sigma_1 =  \sum_{i}  p'_i   P_{i}  \ket{\psi_{c}}\bra{\psi_{c}})   P_{i}  
  \otimes    \ket{ACC} \bra{ACC}  + \sum_{j} p''_j P_{j}  \ket{\psi_{c}}\bra{\psi_{c}}   P_{j}  
  \otimes    \ket{REJ} \bra{REJ}
\end{eqnarray}

where
\begin{itemize}
\item $P_{i}$ ($P_{j}$) range over all general Pauli operators on the output system $O$.
\item $p'_i,p''_i$ are probabilities with $\sum_{i} (p'_i + p''_i)  = \sum_{k,u \in \mathcal{S}_1,v} | a_{k,u,v} |^2$ .
\item $\ket{ACC} \bra{ACC} $ is the state where all trap qubits are $\ket{0}\bra{0}$ and $\ket{REJ} \bra{REJ}$ denotes any  state that at least one trap is $\ket{1}\bra{1}$.
\item $\ket{\psi_{c}}\bra{\psi_{c}}$ is the correct state for the output system $O$, which crucially is placed on a fixed position which does not depend on the selection of the trap positions $(\boldsymbol{t})$ (since the position of $O'$ was also fixed due to the structure of the gadget and the decoding circuit is fixed).
\end{itemize}

We examine the rest of the terms of the summation in the state in Equation \ref{a3} (where $u \in \mathcal{S}_2$ means that $P_{u}$ contains $> d$ Pauli elements with a Pauli $X$ component).

Let $P_{\perp}$ be the projection to the orthogonal space to the correct output state:
\begin{equation}
P_{\perp}=I - \ket{\psi_{c}}\bra{\psi_{c}}
\end{equation}

We calculate the `bad' probability $p_{\text{bad}}$ the state collapses to the `incorrect' subspace and no trap is  activated.

\begin{eqnarray}
p_{\text{bad}} = Tr( \sum_{\boldsymbol{t}} p(\boldsymbol{t}) P_{\perp} \bigotimes_{i}  \ket{0}_{t_i} \bra{0}_{t_i} 
 (\sum_{\boldsymbol{b}}   \sum_{k,u \in \mathcal{S}_2,v}  |a_{k,u,v}|^2   C'_{O'} \mathcal{D} (\bra{\boldsymbol{b}'} P_{u|{i: \forall j, i \neq t_j}}  \otimes P_{v|O''} \nonumber \\ (   \mathcal{P}^{(3)} (  \ket{M^{(3)}}\bra{M^{(3)}}  ) \mathcal{P}^{(3)\dagger}  ) 
      P_{u|{i: \forall j, i \neq t_j}} \otimes P_{v|O''} \ket{\boldsymbol{b}'} ) C'^{\dagger}_{O'}  \bigotimes_{i}   P_{u|{t_i}} \ket{0}_{t_i} \bra{0}_{t_i} P_{u|{t_i}} )
\end{eqnarray}

Tracing out $P_{\perp}  (\sum_{\boldsymbol{b}} C'_{O'} \mathcal{D} (\bra{\boldsymbol{b}'} P_{u|{i: \forall j, i \neq t_j}}  \otimes P_{v|O''}  (   \mathcal{P}^{(3)} (  \ket{M^{(3)}}\bra{M^{(3)}}  ) \mathcal{P}^{(3)\dagger}  ) 
      P_{u|{i: \forall j, i \neq t_j}} \otimes P_{v|O''} \ket{\boldsymbol{b}'} ) C'^{\dagger}_{O'} $:
      
\begin{eqnarray}
\leq Tr(   \sum_{\boldsymbol{t}} p(\boldsymbol{t}) \ket{0}_{t_i} \bra{0}_{t_i}   \sum_{k,u \in \mathcal{S}_2,v}  |a_{k,u,v}|^2   \bigotimes_{i}   P_{u|{t_i}} \ket{0}_{t_i} \bra{0}_{t_i} P_{u|{t_i}} )
\end{eqnarray}

Or:

\begin{eqnarray}
= \prod_{i} \sum_{t_i} p(t_i)   \sum_{k,u\in \mathcal{S}_2,v} | a_{k,u,v} |^2 | \bra{0}_{t_i} P_{u|t_i} \ket{0}_{t_i} |^2 
\end{eqnarray}

We can divide the measured positions into $|\boldsymbol{t}|$ subsets $S_{\gamma}$ where each trap position $t_{\gamma}$ is chosen uniformly at random between the $|S_{t_{\gamma}}|=1/c'$ positions (revealing this information can only deteriorate the security, so the bound will be true also for the previous case):

\begin{eqnarray}
= \prod_{\gamma =1 }^{|\boldsymbol{t}|} \sum_{t_{\gamma} \in S_{\gamma}} p(t_{\gamma})   \sum_{k,u\in \mathcal{S}_2,v} | a_{k,u,v} |^2 | \bra{0}_{t_{\gamma}} P_{u|t_{\gamma}} \ket{0}_{t_{\gamma}} |^2 
\end{eqnarray}

Denoting by $w_{u,\gamma}$ the number of positions such that $ P_{u|t_{\gamma}}$ (over all $1/c'$ values that the particular $t_{\gamma}$ can take)  is either Pauli $X$ or $Y$:

\begin{eqnarray}
=  \prod_{\gamma =1 }^{|\boldsymbol{t}|}  c'  \sum_{k,u\in \mathcal{S}_2,v} | a_{k,u,v} |^2 (\frac{1}{c'} - w_{u,\gamma}) \nonumber \\
\leq  \prod_{\gamma =1 }^{|\boldsymbol{t}|}    \sum_{k,u\in \mathcal{S}_2,v} | a_{k,u,v} |^2 (1 - c')^{w_{u,\gamma}}
\end{eqnarray}

where the last inequality comes from the fact that  $w_{u,\gamma}$ is  non-negative integer.

\begin{eqnarray}
=      \sum_{k,u\in \mathcal{S}_2,v} | a_{k,u,v} |^2 (1 - c')^{\sum_{\gamma =1 }^{|\boldsymbol{t}|} w_{u,\gamma}} \nonumber \\
\leq  \sum_{k,u\in \mathcal{S}_2,v} | a_{k,u,v} |^2 (1 - c')^{d}
\end{eqnarray}

where the last inequality comes from the fact that since $u\in \mathcal{S}_2$, it follows that the total footprint $\sum_{\gamma =1 }^{|\boldsymbol{t}|} w_{u,\gamma}$ of attack $P_u$ is $> d$.

Therefore state  can be written as:

\begin{eqnarray}
\sigma_2 \approx_{ \sqrt{ (1 - c')^d}} p (\sum_{k,u\in \mathcal{S}_2,v} | a_{k,u,v} |^2) \ket{\psi_{c}}\bra{\psi_{c}} \otimes \ket{ACC} \bra{ACC} \nonumber \\
+ (1-p) (\sum_{k,u\in \mathcal{S}_2,v} | a_{k,u,v} |^2) \rho \otimes \ket{REJ} \bra{REJ}
\end{eqnarray}

for some probability $p$ and density matrix $\rho$. Or,

\begin{eqnarray}
\sigma_2 \approx_{ \sqrt{ (1 - c')^d}} p_1   \ket{\psi_{c}}\bra{\psi_{c}} \otimes \ket{ACC} \bra{ACC} 
+ p_2  \rho \otimes \ket{REJ} \bra{REJ}
\end{eqnarray}

where 

\begin{itemize}
\item $p_1+p_2 = \sum_{k,u\in \mathcal{S}_2,v} | a_{k,u,v} |^2$
\item $\rho$ is a density matrix.
\end{itemize}

Summing the terms $\sigma_1$ and $\sigma_2$, Theorem \ref{theorem1}  is satisfied with $\epsilon = \sqrt{ (1 - c')^d}$.

\section{Proof of Theorem \ref{theorem2}} \label{app_b}

\begin{figure}[h!]
  \centering
      \includegraphics[width=1\textwidth]{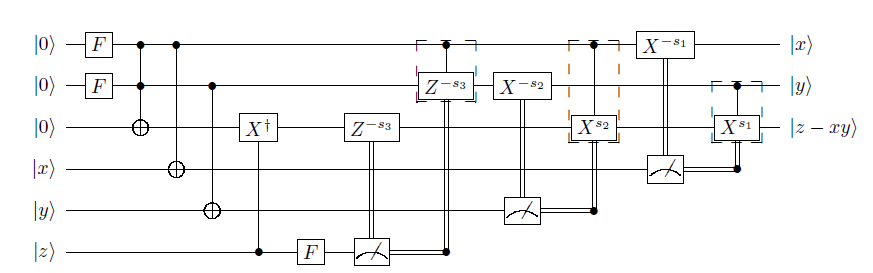}
  \caption{Implementation of Toffoli gate with Toffoli state, Clifford operators and Pauli $Z$ measurements. The corrections (which have to be performed by Bob depending on his communication with Alice) are the Clifford operators in the dashed boxes.} \label{figure_t}
\end{figure}

For any run of Protocol \ref{prot:hybrid_ver}, for any choice of Alice's input computation (therefore for any choice of sign keys used in the ABE protocol phase), the output that Alice has at the end of the protocol before applying the ABE detection procedure, averaged over all random parameters except the sign key, is: 

\begin{eqnarray}
\rho_{out}=Tr_{B,\mathcal{M},\Delta,D,\mathcal{L}} (\sum_{\boldsymbol{b},\boldsymbol{\widetilde{b}}}  \sum_{\nu} p(\nu)   C'' (   \ket{\boldsymbol{b}}\bra{\boldsymbol{b}} \otimes \ket{\boldsymbol{\widetilde{b}}}\bra{\boldsymbol{\widetilde{b}}} U \mathcal{C} \mathcal{D}  (   F_{N-n'} R_{N-n'}(.) \ldots F_1 R_1(.) E_G ( \nonumber \\ \ket{M}\bra{M} 
 \otimes  \ket{\boldsymbol{\delta}}\bra{\boldsymbol{\delta}} ) E_G R_1(.)^{\dagger}  F_1^{\dagger} \ldots R_{N-n'}(.)^{\dagger} F_{N-n'}^{\dagger}   \otimes \ket{0}\bra{0}^{\otimes B} \otimes \ket{\widetilde{\boldsymbol{r}}}\bra{\widetilde{\boldsymbol{r}}})
   \mathcal{D}^{\dagger} \mathcal{C}^{\dagger} U^{\dagger} \ket{\boldsymbol{b}}\bra{\boldsymbol{b}} \otimes \ket{\boldsymbol{\widetilde{b}}}\bra{\boldsymbol{\widetilde{b}}}  ) C'' )
\end{eqnarray}

where:

\begin{itemize}
\item $R_i(\delta_i)$ corresponds the diagonal generalized Clifford operation $T^{c_i} S^{b_i} Z^{a_i}$ controlled by the state of vector $\delta_i$.
\item $F_i$ is the Fourier gate applied on qubit $i$.
\item $\mathcal{D}$ is the Clifford unitary used to decode the output $O'$ of the FK protocol phase which is encoded by the QECC used by the FK protocol. The decoding is applied by Bob in this case and the decoded state is contained in subsystem $O$ of $O'$.
\item $\ket{\boldsymbol{\widetilde{b}}}\bra{\boldsymbol{\widetilde{b}}}$  implement the measurements for the Toffoli gates at the ABE protocol phase and apply on some of the qubits of system $O$ (see Figure \ref{figure_t}).
\item $\widetilde{\boldsymbol{r}}(\boldsymbol{r},\widetilde{\boldsymbol{b}})$ is the classical bit string (system  $\mathcal{L}$) that Alice sends to Bob for the corrections of the Toffoli part of the ABE protocol.
\item $\mathcal{C}(\widetilde{\boldsymbol{r}})$ is the Clifford part of the ABE protocol phase that implements the public polynomial-QECC logical circuit on system $O$.
\item $C''(\boldsymbol{r},\boldsymbol{b})$ contains the Pauli correction $C''_{O'}$ that Alice performs to the final returned system, that takes into account the updates of the keys due to the application of circuits $\mathcal{D}$ and $\mathcal{C}$. $C''$ also contains the corrections to trap measurement results.
\item $U$ represents global Bob's attack that in this case applies also to system $\mathcal{L}$. 
\end{itemize}

Alice's output can be rewritten, by decomposing the attack to the Pauli basis, tracing out system $B$ and applying the honest computation on system $T$ to separate it, as:

\begin{eqnarray}
\rho_{out} =Tr_{\mathcal{M},\Delta,D,\mathcal{L}} (\sum_{\boldsymbol{b},\boldsymbol{\widetilde{b}}}  \sum_{\nu} p(\nu) \sum_{k,u,v,u',v'}  a_{k,u,v} a_{k,u',v'}^*   C''_{O'}  (   \ket{\boldsymbol{b}'}\bra{\boldsymbol{b}'}  \otimes \ket{\boldsymbol{\widetilde{b}}}\bra{\boldsymbol{\widetilde{b}}} P_{u|{i: \forall j, i \neq t_j}}  \otimes P_{v}  \nonumber \\ \mathcal{C} \mathcal{D}(   \mathcal{P}' (  \ket{M'}\bra{M'} 
 \otimes  \ket{\boldsymbol{\delta}}\bra{\boldsymbol{\delta}} ) \mathcal{P}'^{\dagger}  ) \mathcal{D}^{\dagger} \mathcal{C}^{\dagger} 
      P_{u'|{i: \forall j, i \neq t_j}} \otimes P_{v'} \ket{\boldsymbol{b}'}\bra{\boldsymbol{b}'}  \otimes \ket{\boldsymbol{\widetilde{b}}}\bra{\boldsymbol{\widetilde{b}}}  ) C''^{\dagger}_{O'} \nonumber \\ \bigotimes_{i}   \ket{b_{t_i}+r_{t_i}}\bra{b_{t_i}} P_{u|{t_i}} \ket{r_{t_i}} \bra{r_{t_i}} P_{u'|{t_i}} \ket{b_{t_i}}\bra{b_{t_i}+r_{t_i}}))
\end{eqnarray}

where
\begin{itemize}
\item $ a_{k,u,v}$ are complex numbers, with $\sum_{k,u,v} |a_{k,u,v}|^2 = 1$.
\item $P_{u}$ (and $P_{u'}$) ranges over all general $d$-level Pauli operators and applies on the system that is trapified.
\item $P_{v}$ (and $P_{v'}$) ranges over all general $d$-level Pauli operators and applies on the system that is not trapified: $O'' \times \Delta \times \mathcal{L}$.
\item $P_{u|i}$ is the $i$-th generalized Pauli element of $P_{u}$ (similarly for $P_{v}$).
\item $\boldsymbol{b}'$ is the vector that is generated from $\boldsymbol{b}$ by removing elements $\{b_{t_i} \}$. 
\end{itemize}

\hskip 0.1cm

We notice that we can extract a random logical Pauli $Z^{r'_1} \otimes Z^{r'_2} \otimes Z^{r'_3}$ from state $\ket{000}$ that is used to prepare the Toffoli gate (Figure \ref{figure_t}) and notice that the  dependences on $(r'_1,r'_2,r'_3)$ cancel in all systems except system $\mathcal{L}$. Then, summing over $(r'_1,r'_2,r'_3)$ gives the maximally mixed state for system $\mathcal{L}$ and thus taking the trace of it cancels all cross terms of the corresponding attack. The rest of the arguments follow the corresponding part of the proof of Theorem \ref{theorem1}.

Therefore state $\rho_{out}$ can be rewritten to eliminate all Pauli cross terms from the attack  of Bob that are different.

\begin{eqnarray}
\rho_{out} = \sum_{\boldsymbol{b},\boldsymbol{\widetilde{b}}}  \sum_{\boldsymbol{t}} p(\boldsymbol{t}) \sum_{k,u,v}  |a_{k,u,v}|^2   C^{(3)}_{O'}  ( \bra{\boldsymbol{b}'} \otimes \ket{\boldsymbol{\widetilde{b}}}\bra{\boldsymbol{\widetilde{b}}} P_{u|{i: \forall j, i \neq t_j}}  \otimes P_{v|O''}  \nonumber \\ (  \mathcal{C} \mathcal{D}( \mathcal{P}^{(3)} (  \ket{M^{(3)}}\bra{M^{(3)}}  ) \mathcal{P}^{(3)\dagger}) \mathcal{D}^{\dagger} \mathcal{C}^{\dagger}  ) 
      P_{u|{i: \forall j, i \neq t_j}} \otimes P_{v|O''} \ket{\boldsymbol{b}'}  \otimes \ket{\boldsymbol{\widetilde{b}}}\bra{\boldsymbol{\widetilde{b}}} ) C^{(3)\dagger}_{O'}  \bigotimes_{i}   P_{u|{t_i}} \ket{0} \bra{0} P_{u|{t_i}} 
\end{eqnarray}

where

\begin{itemize}
\item $\ket{M^{(3)}}$ is the tensor product of $\ket{+_0}$ for all qudits of system $\mathcal{M} \times O'$.
\item $\mathcal{P}^{(3)}$ is the unitary that contains $E_G'$ for $\mathcal{M} \times O'$ and $F_{i} R_{i}(\phi'_{i}(b_{f^{-1}(i)},b_{f^{-1}} (j \sim i, j \neq f(i))) $ operators for all qudits of system $\mathcal{M}$.
\item $C^{(3)}_{O'}$ are the corrections on the output of the ABE protocol phase (after eliminating the $r$'s) and depend on  $\boldsymbol{b}'$ and $\boldsymbol{\widetilde{b}}$ . 
\end{itemize}

Restricting the summation to $u \in \mathcal{S}_1$ where $\mathcal{S}_1$ is the set of general $d$-level Pauli operators that contain $\leq d_1$ elements which have a generalized Pauli $X$ component (where $d_1$ is the number of correctable errors of the code)
and applying the same steps as in the corresponding part of the proof of Theorem \ref{theorem1}, the state becomes:

\begin{eqnarray}
\sigma_1 =\sum_{\boldsymbol{b}}  \sum_{\boldsymbol{t}} p(\boldsymbol{t}) \sum_{k,u \in \mathcal{S}_1,v}  |a_{k,u,v}|^2      C^{(3)}_{O'} P''_{v_1} P'_{v_2}  (   C^{(3)}_{O'} \ket{\psi_{c}'}\bra{\psi_{c}'} C^{(3)}_{O'} ) P'_{v_2} P''_{v_1}  C^{(3)}_{O'}   )   \bigotimes_{i}   P_{u|{t_i}} \ket{0} \bra{0} P_{u|{t_i}} 
\end{eqnarray}

The fact that the graph is partitioned into sub-graphs, to encode the different logical states to be used in the ABE protocol phase does not change the above statement because a global attack with footprint $\leq d_1$ will necessarily have footprint $\leq d_1$ in each sub-graph. We can also eliminate $C^{(3)}_{O'}$ by commuting with the Pauli attack operators.

By separating the terms that leave all traps untouched and the rest we get: 

\begin{eqnarray}
\sigma_1 =  \sum_{i}  p'_i   P_{i} \ket{\psi_{c}'}\bra{\psi_{c}'}   P_{i}  
  \otimes    \ket{ACC_1} \bra{ACC_1}  + \sum_{j} p''_j P_{j} \ket{\psi_{c}'}\bra{\psi_{c}'}    P_{j}  
  \otimes    \ket{REJ_1} \bra{REJ_1}
\end{eqnarray}

where
\begin{itemize}
\item $P_{i}$ ($P_{j}$) range over all general $d$-level Pauli operators on the final returned system.
\item $p'_i$ , $p''_i$ are probabilities with $\sum_{i} (p'_i + p''_i)  = \sum_{k,u \in \mathcal{S}_1,v} | a_{k,u,v} |^2$ .
\item $\ket{ACC_1} \bra{ACC_1} $ is the state where all trap qubits are $\ket{0}\bra{0}$ and $\ket{REJ_1} \bra{REJ_1}$ denotes any  state that at least one trap is $\ket{1}\bra{1}$.
\item $\ket{\psi_{c}'}\bra{\psi_{c}'}$ is the state, encoded by the signed polynomial QECC, after the correct computation of the ABE protocol phase is performed and undoing the Pauli key.
\end{itemize}

The rest of the terms of the summation in the state $\rho_{out}$ ( $u \in \mathcal{S}_2$ ) are considered. Let $P_{\perp}$ be the projection to the orthogonal space to the correct output state:
\begin{equation}
P_{\perp}=I - \ket{\psi_{c}'}\bra{\psi_{c}'}
\end{equation}

We calculate the `bad' probability $p_{\text{bad}}$ the state collapses to the `incorrect' subspace and no trap is  activated.

\begin{eqnarray}
p_{\text{bad}}= Tr( \sum_{\boldsymbol{b},\boldsymbol{\widetilde{b}}}  \sum_{\boldsymbol{t}} p(\boldsymbol{t}) P_{\perp} \bigotimes_{i}  \ket{0}_{t_i} \bra{0}_{t_i}  \sum_{k,u \in \mathcal{S}_2,v}  |a_{k,u,v}|^2   C^{(3)}_{O'}  ( \bra{\boldsymbol{b}'} \otimes \ket{\boldsymbol{\widetilde{b}}}\bra{\boldsymbol{\widetilde{b}}} P_{u|{i: \forall j, i \neq t_j}}  \otimes P_{v|O''}  \nonumber \\ (  \mathcal{C} \mathcal{D}( \mathcal{P}^{(3)} (  \ket{M^{(3)}}\bra{M^{(3)}}  ) \mathcal{P}^{(3)\dagger})  \mathcal{D}^{\dagger} \mathcal{C}^{\dagger}  ) 
      P_{u|{i: \forall j, i \neq t_j}} \otimes P_{v|O''} \ket{\boldsymbol{b}'}  \otimes \ket{\boldsymbol{\widetilde{b}}}\bra{\boldsymbol{\widetilde{b}}} ) C^{(3)\dagger}_{O'}  \bigotimes_{i}   P_{u|{t_i}} \ket{0} \bra{0} P_{u|{t_i}} )
\end{eqnarray}

Tracing out everything except the trap system:

\begin{eqnarray}
\leq Tr(   \sum_{\boldsymbol{t}} p(\boldsymbol{t}) \ket{0}_{t_i} \bra{0}_{t_i}   \sum_{k,u \in \mathcal{S}_2,v}  |a_{k,u,v}|^2   \bigotimes_{i}   P_{u|{t_i}} \ket{0}_{t_i} \bra{0}_{t_i} P_{u|{t_i}} )
\end{eqnarray}

The rest is the same to the corresponding part in the proof of Theorem \ref{theorem1} giving the final state:

\begin{eqnarray}
\sigma_2 \approx_{ \sqrt{ (1 - c')^{d_1}}} p_1   \ket{\psi_{c}'}\bra{\psi_{c}'}  \otimes \ket{ACC_1} \bra{ACC_1} 
+ p_2  \rho \otimes \ket{REJ_1} \bra{REJ_1}
\end{eqnarray}

where $p_1+p_2 = \sum_{k,u\in \mathcal{S}_2,v} | a_{k,u,v} |^2$ and $\rho$ is a density matrix.

By summing the states $\sigma_1$ and $\sigma_2$ and by contractiveness of the ABE protocol detection procedure that Alice applies to check if the state is in the valid space of the signed polynomial QECC, Theorem \ref{theorem2} is satisfied with $\epsilon_1 =  \sqrt{ (1 - c')^{d_1}}$ and  $\epsilon_2 =   (1/2)^{d_2}$.

\end{document}